\documentclass[review,times]{elsarticle}
\usepackage[utf8]{inputenc}

\usepackage[colorlinks=true,linkcolor=black, citecolor=blue, urlcolor=blue]{hyperref}

\usepackage[dvipsnames]{xcolor}
\usepackage{xspace,soul}
\usepackage{makecell}
\usepackage{rotating}
\usepackage{ulem}
\usepackage{float}

\usepackage{setspace}
\usepackage{orcidlink}
\usepackage{amssymb}
\usepackage{amsmath}

\newcommand{\gobble}[1]{}
\makeatletter
\newcommand\newref[1]{#1\def\@currentlabel{#1}}
\makeatother 
\begin{document}
\begin{frontmatter}

%
%
\title{Systematic effects on a Compton polarimeter at the focus of an X-ray mirror}
%
%
\author[Osaka]{M.\,Aoyagi}
\author[WUSTL,WUSTL-McDonnel]{R.~G.\,Bose}
\author[WUSTL,WUSTL-McDonnel]{S.\,Chun\orcidlink{0009-0002-2488-5272}}
\author[WUSTL,WUSTL-McDonnel]{E.\,Gau\orcidlink{0000-0002-5250-2710}}
\author[WUSTL,WUSTL-McDonnel]{K.\,Hu\orcidlink{0000-0002-9705-7948}}
\author[Osaka]{K.\,Ishiwata\orcidlink{0000-0003-3176-1438}}
\author[KTH1,KTH2]{N.~K.\,Iyer\orcidlink{0000-0001-6134-8105}}
\author[UNH]{F.\,Kislat\orcidlink{0000-0001-7477-0380}}
\author[KTH1,KTH2]{M.\,Kiss\orcidlink{0000-0001-5191-9306}\corref{cor1}}
\ead{mozsi@kth.se}
\author[KTH1,KTH2]{K.\,Klepper}
\author[WUSTL,WUSTL-McDonnel]{H.\,Krawczynski\orcidlink{0000-0002-1084-6507}}
\author[WUSTL,WUSTL-McDonnel]{L.\,Lisalda\orcidlink{0000-0002-5202-1642}} 
\author[ISAS]{Y.\,Maeda\orcidlink{0000-0002-9099-5755}}
\author[KTH1,KTH2]{F.\,af~Malmborg\orcidlink{0009-0007-2274-2055}}
\author[Osaka,Osaka-Forefront]{H.\,Matsumoto}
\author[TokyoMet]{A.\,Miyamoto}
\author[OIST]{T.\,Miyazawa\orcidlink{0000-0002-6068-6337}}
\author[KTH1,KTH2]{M.\,Pearce\orcidlink{0000-0001-7011-7229}}
\author[WUSTL,WUSTL-McDonnel]{B.~F.\,Rauch\orcidlink{0000-0002-1452-4142}}
\author[WUSTL,WUSTL-McDonnel]{N.\,Rodriguez~Cavero\orcidlink{0000-0001-5256-0278}}
\author[UNH]{S.\,Spooner\orcidlink{0000-0003-0710-8893}}
\author[Hiroshima]{H.\,Takahashi\orcidlink{0000-0001-6314-5897}}
\author[TokyoSci]{Y.\,Uchida\orcidlink{0000-0002-7962-4136}}
\author[WUSTL,WUSTL-McDonnel,Arizona]{A.~T.\,West\orcidlink{0000-0002-5471-4709}}
\author[UNH]{K.\,Wimalasena\orcidlink{0000-0003-4434-1766}}
\author[Osaka]{M.\,Yoshimoto\orcidlink{0009-0005-0819-0819}}
\address[Osaka]{Osaka University, Department of Earth and Space Science, Graduate School of Science, and Project Research Center for Fundamental Sciences, 1-1 Machikaneyama-cho, Toyonaka, Osaka 560-0043, Japan}
\address[WUSTL]{Physics Department, Washington University in St. Louis, 1 Brookings Drive, CB 1105, St. Louis, MO 63130, USA}
\address[WUSTL-McDonnel]{McDonnell Center for the Space Sciences, Washington University in St. Louis, 1 Brookings Drive, CB 1105, St. Louis, MO 63130, USA}
\address[KTH1]{KTH Royal Institute of Technology, Department of Physics, 106 91 Stockholm, Sweden}
\address[KTH2]{The Oskar Klein Centre for Cosmoparticle Physics, AlbaNova University Center, 106 91 Stockholm, Sweden}
\address[UNH]{University of New Hampshire, Department of Physics and Astronomy, and Space Science Center, Morse Hall, 8 College Road, Durham, NH 03824, USA}
\address[ISAS]{Institute of Space and Astronautical Science, Japan Aerospace Exploration Agency, 3-1-1 Yoshinodai, Sagamihara, Kanagawa 229-8510, Japan}
\address[Osaka-Forefront]{Forefront Research Center, Osaka University}
\address[TokyoMet]{Department of Physics, Tokyo Metropolitan University, 1-1 Minami-Osawa, Hachioji, Tokyo 192-0397, Japan}
\address[OIST]{Okinawa Institute of Science and Technology Graduate University 1919-1 Tancha, Onna-son, Kunigami-gun Okinawa, Japan 904-0495}
\address[Hiroshima]{Hiroshima University, 1-3-1 Kagamiyama, Higashi-Hiroshima, Hiroshima 739-8526, Japan}
\address[TokyoSci]{Tokyo University of Science 2641 Yamazaki, Noda, Chiba 278-8510, Japan}
\address[Arizona]{Currently at University of Arizona, Department of Astronomy, Steward Observatory, 933 North Cherry Avenue, Tucson, AZ 85721-0065, USA}
\cortext[cor1]{Corresponding author at: KTH Royal Institute of Technology, Department of Physics, 106 91 Stockholm, Sweden}
%
%
\begin{abstract}
{\it XL-Calibur} is a balloon-borne Compton polarimeter for X-rays in the $\sim$15--80~keV range. Using an X-ray mirror with a 12~m focal length for collecting photons onto a beryllium scattering rod surrounded by CZT detectors, a minimum-detectable polarization as low as $\sim$3\% is expected during a 24-hour on-target observation of a 1~Crab source at 45$^{\circ}$ elevation.

Systematic effects alter the reconstructed polarization as the mirror focal spot moves across the beryllium scatterer, due to pointing offsets, mechanical misalignment or deformation of the carbon-fiber truss supporting the mirror and the polarimeter. Unaddressed, this can give rise to a spurious polarization signal for an unpolarized flux, or a change in reconstructed polarization fraction and angle for a polarized flux. Using bench-marked Monte-Carlo simulations and an accurate mirror point-spread function characterized at synchrotron beam-lines, systematic effects are quantified, and mitigation strategies discussed. By recalculating the scattering site for a shifted beam, systematic errors can be reduced from several tens of percent to the few-percent level for any shift within the scattering element. The treatment of these systematic effects will be important for any polarimetric instrument where a focused X-ray beam is impinging on a scattering element surrounded by counting detectors.
\end{abstract}

%
\end{frontmatter}
%
\section{Introduction}
\label{Introduction}
X-ray polarimetry in the hard X-ray band~\cite{DelMonte2022} ($\sim$10--100~keV) probes compact celestial sources, such as X-ray binary black holes and accretion- and rotation-powered neutron stars, which cannot be directly imaged~\cite{Chattopadhyay.2021,Krawczynski.2011lxm}. 
Polarized emission arises from source asymmetry caused by magnetic fields or the distribution of radiation/matter.
The linear polarization of hard X-rays can be measured using Compton scattering, described by the Klein-Nishina differential scattering cross-section~\cite{Klein1929}, 
\begin{equation}
\frac{\mathrm{d}\sigma}{\mathrm{d}\Omega} \propto 2 \sin^2 \theta \cos^2 \phi,
\label{eqn:klein-nishina}
\end{equation}
where $\theta$ is the polar photon scattering angle and $\phi$ is the azimuthal angle between the scattering direction and the polarization vector of the incident photon. The distribution of azimuthal scattering angles is described by a modulation curve with 180$^{\circ}$ periodicity: 
\begin{equation}
C(\phi)=N\{1+\mu\cos[2(\phi-\xi)]\},
\label{eqn:modcurve}
\end{equation}
where $\mu$ and $\xi$ are the amplitude and the phase of the modulation curve, respectively, and $N$ is the mean.  
The polarization fraction, $p$~(\%), is defined as $\mu/\mu_{100}$, where the denominator is the modulation amplitude for a 100\% polarized beam. Since X-rays preferentially scatter perpendicular to the polarization (electric field) vector, see Eq.~(\ref{eqn:klein-nishina}), the polarization angle, $\psi$~($^{\circ}$), can be determined from the modulation phase as $\psi=\xi+90^\circ$. 

Scattering-polarimeter designs often combine a low-atomic-number scatterer symmetrically surrounded by high-atomic-number X-ray detectors, allowing the distribution of azimuthal scattering angles to be sampled over the range of polar angles subtended~\cite{DelMonte2022}. 
In the balloon-borne polarimeter {\it XL-Calibur}~\cite{Abarr.2021}, X-rays are focused over 12~m onto a 12~mm diameter beryllium (Be) rod, whence scattered X-rays are detected by pixelated CdZnTe (CZT) detectors arranged around the rod. The assembly is continuously rotated about the center-line of the Be rod to mitigate temporal and spatial variations in detector response. {\it XL-Calibur} is operating in the $\sim$15--80~keV energy range. The lower bound is due to the limited atmospheric transmission at ballooning altitudes of $\sim$40~km, while the upper limit is dictated by the effective area of the X-ray mirror~\cite{Kamogawa.2022}. For typical source spectra, the $\sim$20--40~keV range will be most favorable for polarimetry.

The next {\it XL-Calibur} flight is scheduled for the spring/summer of 2024, from Esrange Space Center in northern Sweden. The science programme for the two primary northern-sky targets, the Crab pulsar and the black-hole binary Cygnus X-1, requires percent-level minimum detectable polarization (MDP)~\cite{Abarr.2021} (see Section~\ref{Modulation response and spectro-polarimetry} and Eq.~(\ref{MDP equation}) for a definition) in order to determine where in the magnetosphere high-energy emission originates, and to constrain the space-time geometry and coronal shapes in the vicinity of the black hole, respectively. 

Equation~(\ref{eqn:modcurve}) implicitly assumes that scatterings take place along the symmetry axis of the scatterer. For {\it XL-Calibur}, the beam from the X-ray mirror will impinge across the end of the Be rod, as characterized by the mirror point-spread function (PSF). Photons that do not scatter from the symmetry axis will both alter the 180$^{\circ}$ polarization signature and introduce additional harmonics in the modulation response if scattering along the symmetry axis is assumed. Additional components are dominated by, but not restricted to, 360$^{\circ}$ periodicity. They are not canceled out by rotating the polarimeter, and their presence in the modulation response acts as a marker of systematic effects influencing the reconstructed polarization.

Systematic effects from an offset beam were first explored for the Stellar X-Ray Polarimeter instrument, which was developed for the (ultimately canceled) Spectrum X-Gamma mission~\cite{Elsner.1990}. A lithium cylinder served as a Thomson-scattering target and xenon-filled imaging proportional counters were used to detect scattered X-rays in the 4--20~keV  range. The effect of an offset beam on {\it X-Calibur} (the predecessor of {\it XL-Calibur}) was studied using a synchrotron beam and Monte-Carlo simulations~\cite{Beilicke.2014}. 
The scatterer comprised a 13~mm diameter plastic-scintillator rod and the configuration of CZT detectors was different from that in {\it XL-Calibur}. Insights into the response systematics inform the data analysis strategy for {\it XL-Calibur}, but are not fully representative for the realistic mirror PSF, since the beam used was circularly symmetric and the polarimeter was not rotating.

A subsequent iteration of {\it X-Calibur} (8~m focal-length telescope), which used a 12~mm diameter Be scattering rod and a CZT arrangement similar to {\it XL-Calibur}, observed the accretion-powered X-ray pulsar GX\,301-2 during an Antarctic balloon-flight in December 2018~\cite{Abarr.2020p1n}. The polarimeter rotated during observations, and the focal-point offset was corrected for when calculating scattering angles, but systematic effects could not be studied in detail as a balloon leak forced an early termination of the flight. 

In this paper, the modulation response of {\it XL-Calibur} is characterized in detail, using Geant4~\cite{Allison.2016,Agostinelli.2003} Monte-Carlo simulations, for a realistic beam incident on the beryllium stick, including the source spectrum, atmospheric attenuation and the measured mirror PSF. The fidelity of the simulation is established by comparing the measured and simulated instrument response to beams of both polarized and unpolarized X-rays\footnote{Polarization is a positive-definite quantity -- attempting to measure a truly unpolarized (randomly polarized) flux will always yield a net positive polarization fraction~\cite{Mikhalev_Pitfalls_2018}.}. Systematic effects of shifting the PSF in arbitrary directions from the geometric center of the Be rod are studied, since the position of the mirror focal point will change during flight, e.g., due to pointing offsets, and thermal or gravitational forces on the structure. Methods to determine the beam offset and correct the modulation response are discussed, along with strategies to improve sensitivity.

Throughout this paper, polarization parameters are determined using modulation curves. Polarimeter data is commonly analyzed using Stokes parameters, as outlined e.g. in~\cite{Kislat.2015}, due to attractive mathematical properties, such as the additive nature of the parameters allowing simple background subtraction, and their more straight-forward (Gaussian) statistical properties. By definition, the linear-polarization Stokes parameters are only sensitive to the 180$^{\circ}$ harmonic. Asymmetries prominent in the modulation curve may remain undetected if Stokes parameters are directly applied, and the Stokes formalism should therefore only be used after inspecting the modulation curve for additional harmonics, or after applying suitable corrections to the scattering angles.

\section{XL-Calibur design}
\label{XL-Calibur design}
\begin{figure}[hbtp]
    \centering
    \includegraphics[width=.95\textwidth]{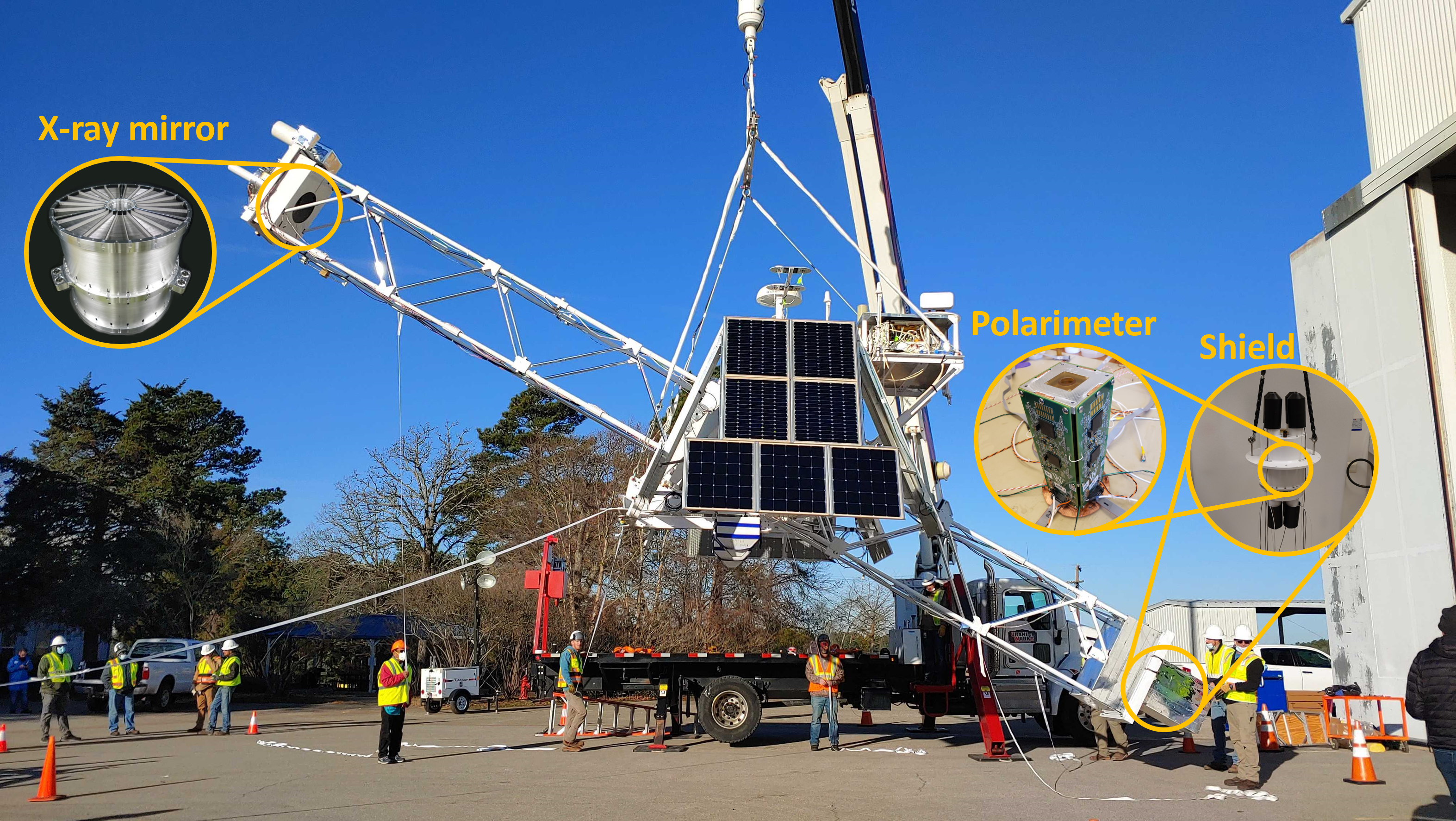}
    \caption{\label{Payload}The {\it XL-Calibur} payload during integration.}
\end{figure}

The {\it XL-Calibur} telescope~\cite{Abarr.2021} comprises a 12~m long rigid truss, with carbon-fiber struts interconnected by aluminum joints. Suspended under a $\sim$10$^6$~m$^3$ stratospheric balloon, the truss can be aimed with arc-second precision using the Wallops Arc-Second Pointer (WASP)~\cite{WASP}.  A 45~cm diameter X-ray mirror is mounted at one end of the truss, with the polarimeter at the other end, as shown in Figure~\ref{Payload}.

The mirror comprises 213 nested aluminum reflector shells with a platinum-carbon multi-layer coating. Shells are arranged in three sections, each spanning 120$^{\circ}$ of the mirror surface. For a 5.9~arcmin field-of-view, the effective area is $\sim$300~cm$^2$ at 20~keV and $\sim$130~cm$^2$ at 40~keV, with a sharp reduction at 78~keV owing to the K~absorption edge of platinum. The half-power diameter (HPD) of the PSF has been measured at the SPring--8 synchrotron facility in Japan as $\sim$2.1~arcmin~\cite{Kamogawa.2022}. In the focal plane of the mirror, at a distance of 12~m, this corresponds to a diameter of $\sim$7.3~mm.

\begin{figure}[hbtp]
    \centering
    \includegraphics[width=.95\textwidth]{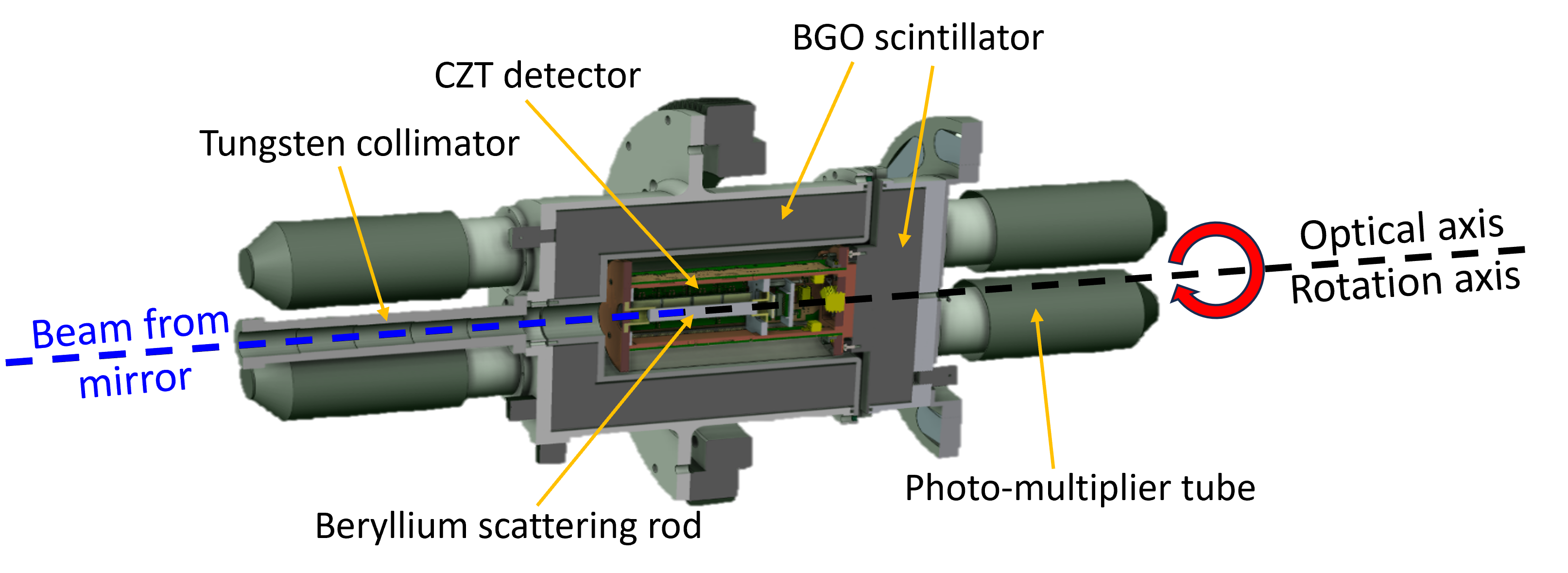}
    \caption{\label{Polarimeter sketch}Overview of the {\it XL-Calibur} shield and polarimeter. The focused X-ray beam from the mirror impinges on a beryllium rod and scattered photons are registered in the surrounding CZT detectors.}
\end{figure}

The polarimeter is housed inside a Bi$_4$Ge$_3$O$_{12}$ (BGO) anticoincidence shield~\cite{Iyer.2023}, as shown in Figure~\ref{Polarimeter sketch}.
Source X-rays focused by the mirror pass through a tungsten collimator and impinge upon a beryllium rod (diameter 12~mm, length 80~mm), surrounded by 20~mm (width) $\times$ 20~mm (height) $\times$ 0.8~mm (thickness) CZT detectors, arranged in a square geometry. A set of 16 CZT detectors, each with 8~$\times$~8 pixels, comprise the four detector walls, for a total of 1024 channels. The CZTs have an energy resolution of $\sim$5.9~keV (FWHM) at 40~keV, enabling spectro-polarimetry. An additional CZT \mbox{({\textquotedblleft}detector~17{\textquotedblright)}} is mounted beneath the beryllium rod, allowing the profile and position of the focused beam to be monitored. This information is supplemented by alignment data from a back-looking camera. Mounted inside the mirror along the optical axis, the camera allows truss deformation to be measured. Positions are determined up to several times per minute, by transferring only offset coordinates based on image-recognition of a keyed circular pattern of LED lights mounted at the polarimeter front~\cite{Abarr.20224ji}.  During observations, the polarimeter/shield assembly continuously rotates about the viewing axis approximately twice per minute.

\section{Simulation and validation}
\label{Simulation and validation}

\begin{figure}[hbtp]
    \centering
    \includegraphics[width=.475\textwidth]{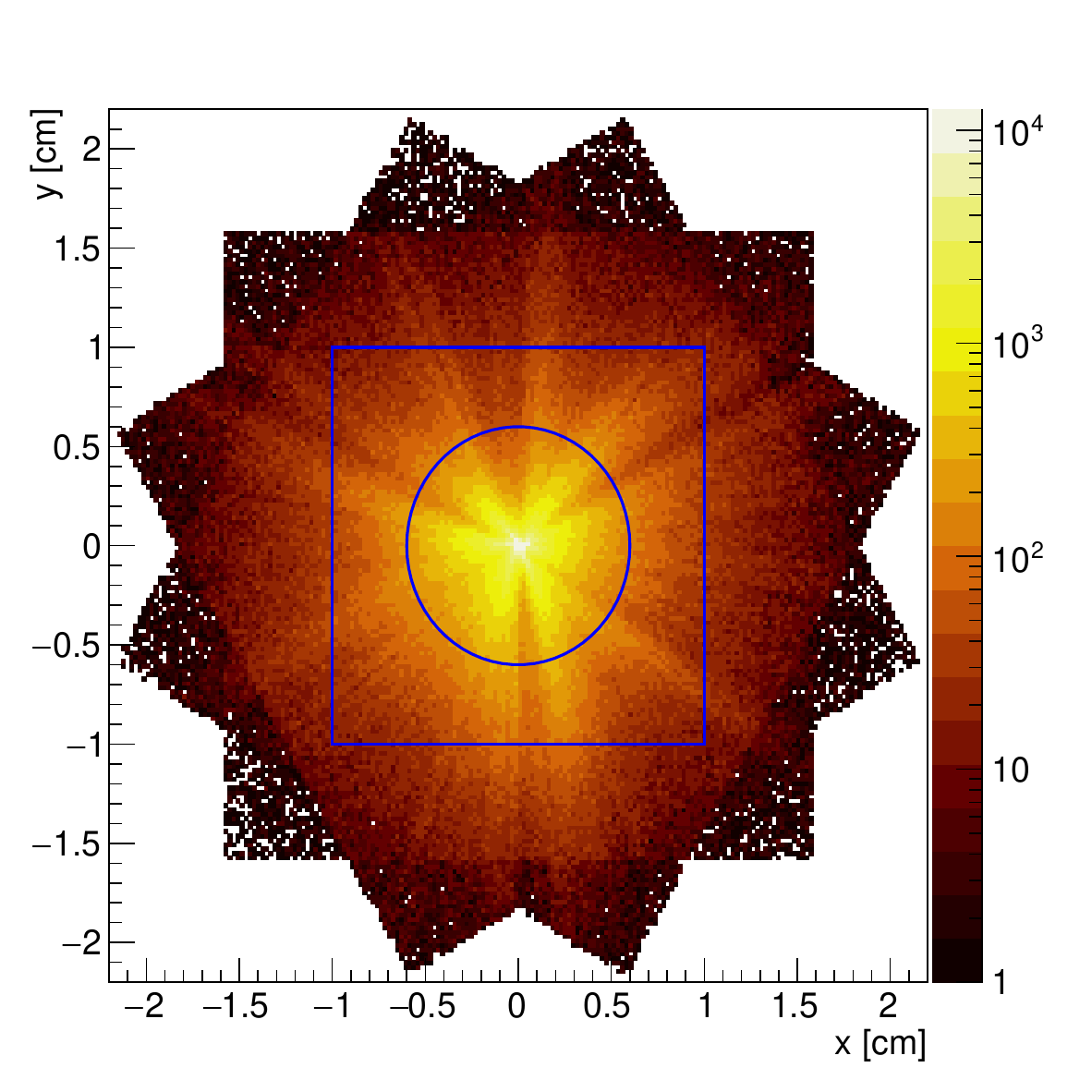}
    \includegraphics[width=.475\textwidth]{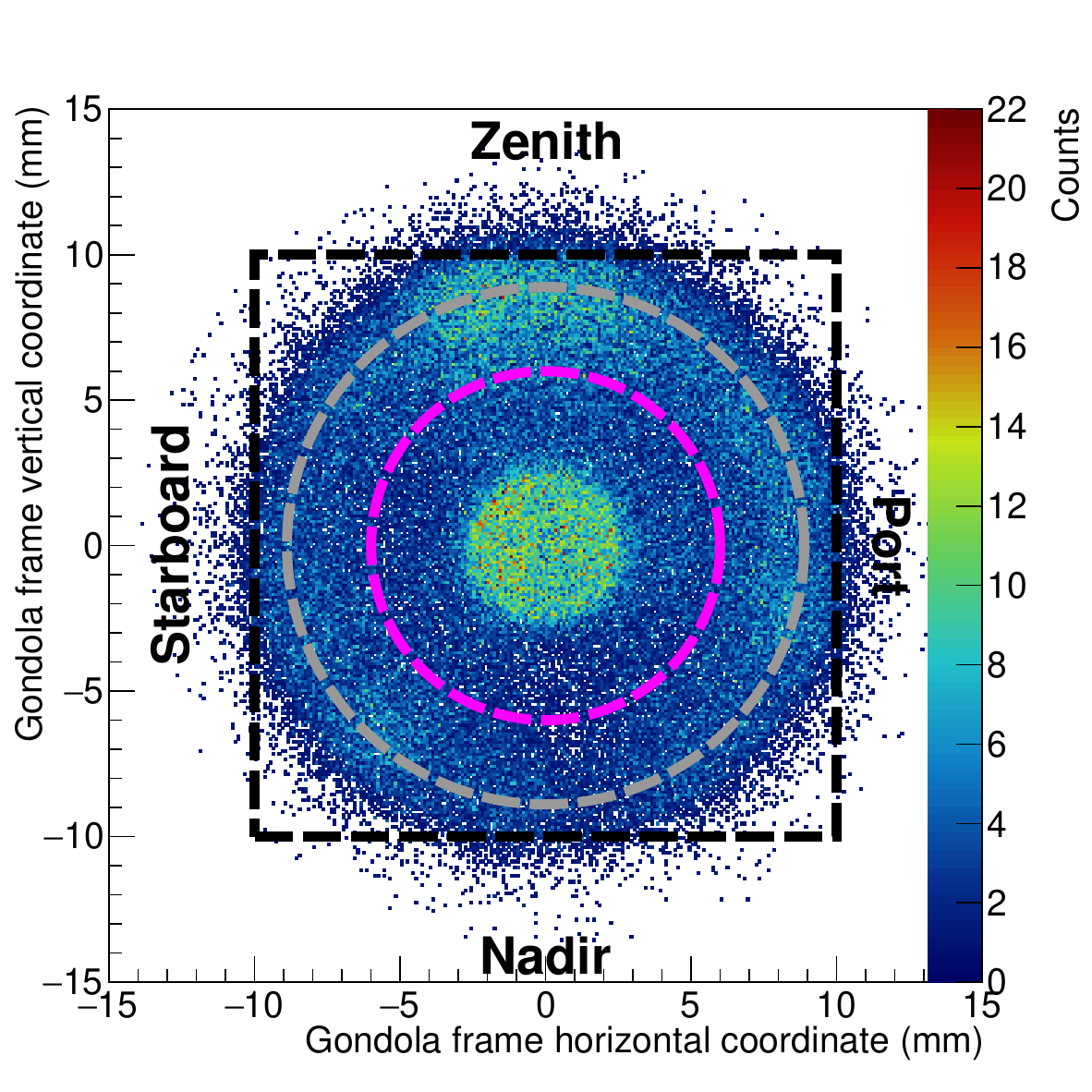}
    \caption{\label{PSF and D17}Left: point-spread function (PSF) of the {\it XL-Calibur} mirror, measured at the SPring--8 synchrotron facility (logarithmic color scale). The twelve-cornered star-shape results from three images, recorded with a square CMOS camera during the beam test~\cite{Kamogawa.2022}, rotated by 120$^{\circ}$ relative to each other, with one image for each mirror section. The blue circle and blue square indicate the beryllium stick and the CZT walls, respectively. Right: the resulting simulated distribution of photon hits in detector~17, deconvolved for the detector roll. For each hit above the detection threshold, the coordinate is randomized within the \mbox{2.5 mm $\times$ 2.5 mm} pixel size, to provide sub-pixel binning. The sharp boundary near the center of the image is from the central \mbox{2 $\times$ 2} pixels detecting most of the hits. Asymmetries arise from the true mirror behavior, with subtle differences between mirror sections. Even though the mirror is perfectly aligned in the simulation, the resulting beam distribution within the beryllium stick (dashed magenta line) is shifted, by $\sim$0.3~mm, approximately in the zenith direction. Dashed black lines represent the detector walls and the size of detector~17 (\mbox{20 mm $\times$ 20 mm}). As the detector rotates, events outside this region can become populated. The tungsten collimator is tapered to a minimum diameter of $\sim$18~mm (dashed gray line) to prevent direct illumination of the CZT walls.}
\end{figure}

A Monte-Carlo simulation has been implemented within the Geant4 framework (version~10.02) to model the polarimeter response. This simulation was previously used for the mission design~\cite{Abarr.2021} and for optimizing the anticoincidence-shield configuration~\cite{Iyer.2023}. For the current study, the mirror point-spread function was also implemented. Source photons are sampled from a given spectrum and propagated with a spatial distribution based on beam-line calibration data~\cite{Kamogawa.2022} for the three mirror sections. Each section was tested at energies 20, 30, 40, 50 and 70~keV, and the simulated PSF uses the measured mirror response at the closest energy. The PSF can be translated across the face of the scattering element to allow the study off-center pointings or misalignments. Figure~\ref{PSF and D17} shows an example of the mirror PSF implemented in the simulation, and the resulting hit-distribution in detector~17, for a power-law spectrum with photon index~2.1. 

To bench-mark the simulations, measurements with a $\sim$1.75~GBq $^{241}$Am source were performed at Esrange, immediately prior to the 2022 {\it XL-Calibur} launch\footnote{For these measurements, a temporary copper collimator was mounted into the polarimeter bore-sight, to prevent the diverging beam from the radioactive source from illuminating the CZT walls directly.}. In one configuration, source photons directly illuminated the beryllium rod, allowing the response to unpolarized radiation to be studied. In a second configuration, source photons were scattered through 90$^{\circ}$ in a dedicated polarizer, resulting in a beam with close to 100\% polarization. Emission lines from the radioactive decay of $^{241}$Am were simulated with their relevant branching ratios, with the most prominent X-ray line at 59.5~keV. Generated source photons were propagated through the simulated experimental setup, with no manual tuning of spectra or photon spatial distributions.

Polarized (unpolarized) measurement data was acquired over 36 hours (1 hour), yielding 954930 (1692873) valid events (hit in one CZT pixel only and no coincident hit in the anticoincidence shield above a 100~keV threshold), for a rate of $\sim$7~Hz ($\sim$500~Hz). The contemporaneous background trigger rate was $\sim$1~Hz. Signal (source exposed) and background (source blocked) observations were interspersed, allowing accurate background subtraction under variable measurement conditions. Based on relative rates, the beam-on/beam-off ratio was taken to be 2:1 (10:1) for the polarized (unpolarized) beam, which ensures the minimization of the uncertainty incurred by the background subtraction~\cite{Kislat.2015}. These rates and exposure times result in a statistical error on the reconstructed modulation amplitude of $<$0.2~percentage points.

\begin{figure}[hbt]
    \centering
    \includegraphics[width=.475\textwidth]{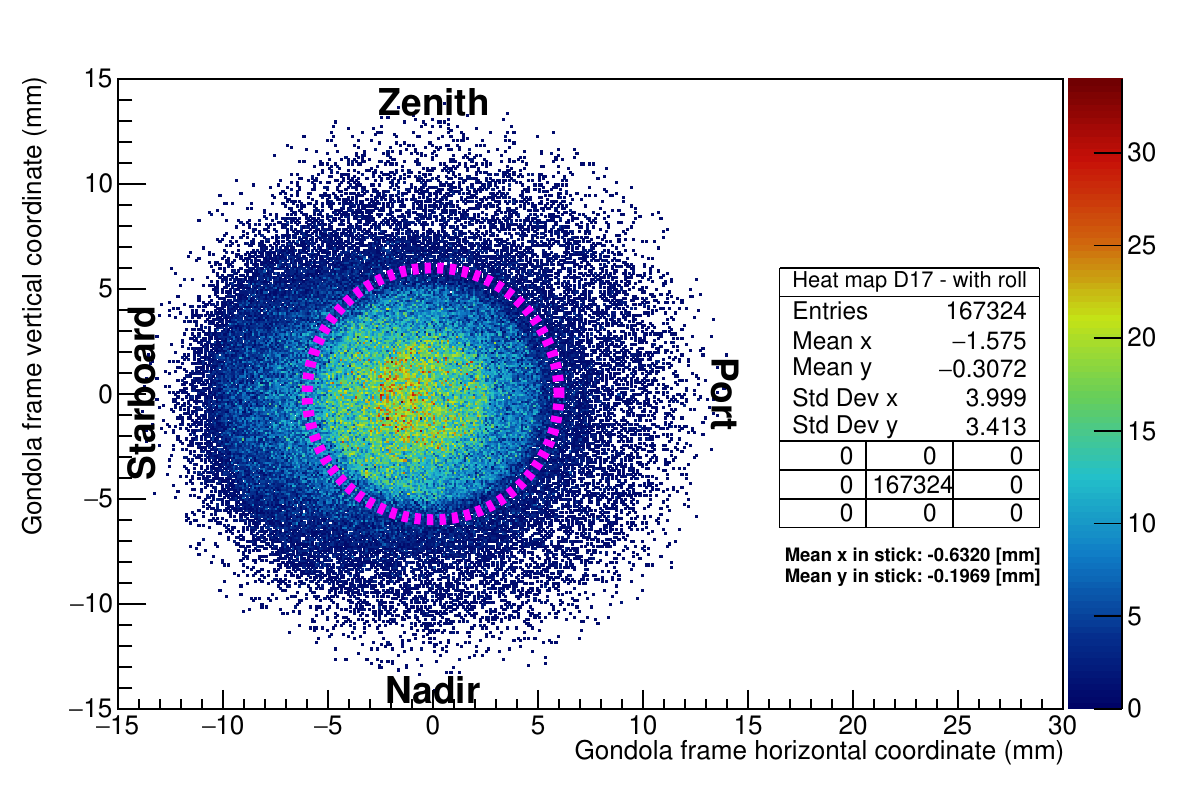}
    \includegraphics[width=.475\textwidth]{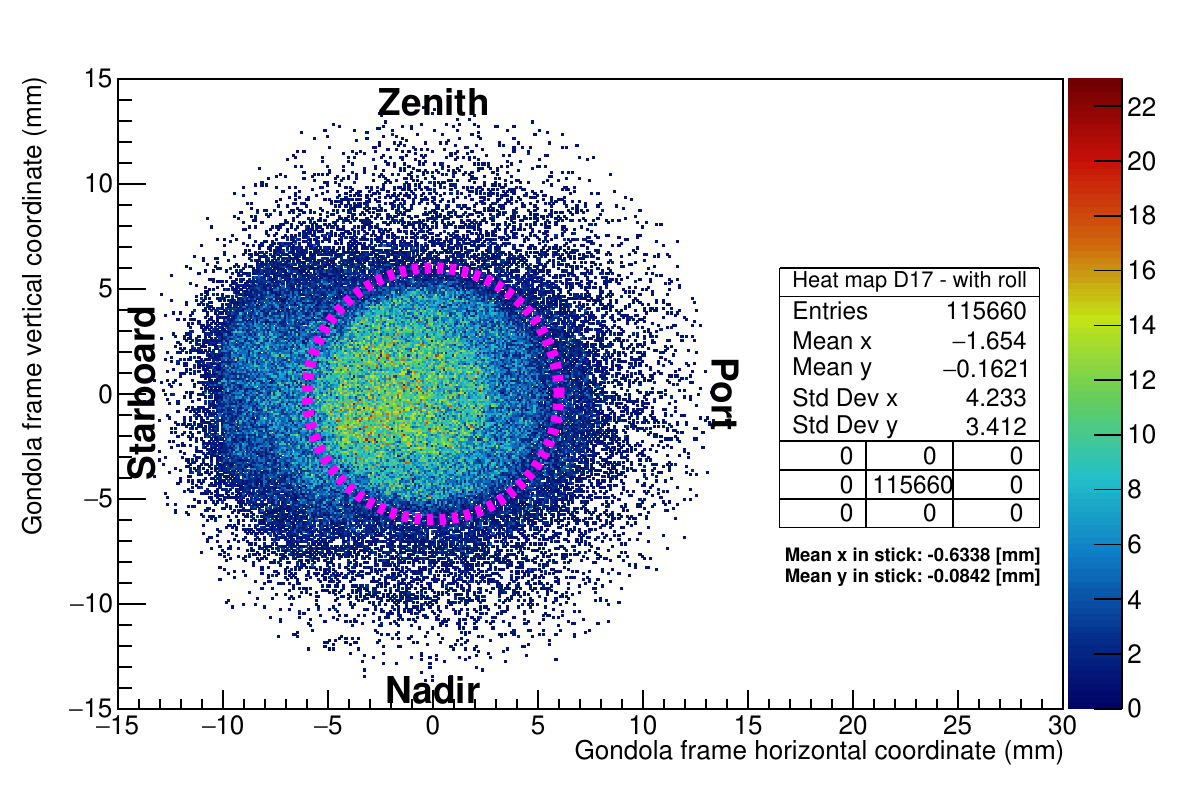}
    \caption{\label{Weekend run - D17 images}Detector-17 hit-maps from the measurement (left) and simulation (right). Mean~x and mean~y values indicate that the beam is hitting the polarimeter off the optical/X-ray/rotation axis, due to a shift and/or a tilt in the radioactive-source alignment. The size of the beryllium scatterer is shown by the magenta circle. Mean~x and mean~y values for events hitting within the radius of the rod have been separately indicated.}
\end{figure}

The modulation curve for the polarized beam is not well-described by Eq.~(\ref{eqn:modcurve}): additional harmonics, such as a 360$^{\circ}$ and a 120$^{\circ}$ component, must be introduced to fit the angular distribution. These components arise from an offset of the beam impinging on the scatterer. To accurately replicate the offset present in the alignment of the experimental setup, the simulated position of the radioactive source geometry was moved iteratively in x and y, minimizing the difference between measured/simulated beam center-point positions, as reconstructed in detector~17 by the arithmetic (mean~x, mean~y). The result is shown in Figure~\ref{Weekend run - D17 images}, where the central positions agree within $\sim$0.1~mm.

\begin{figure}[hbt]
    \centering
    \includegraphics[width=.475\textwidth]{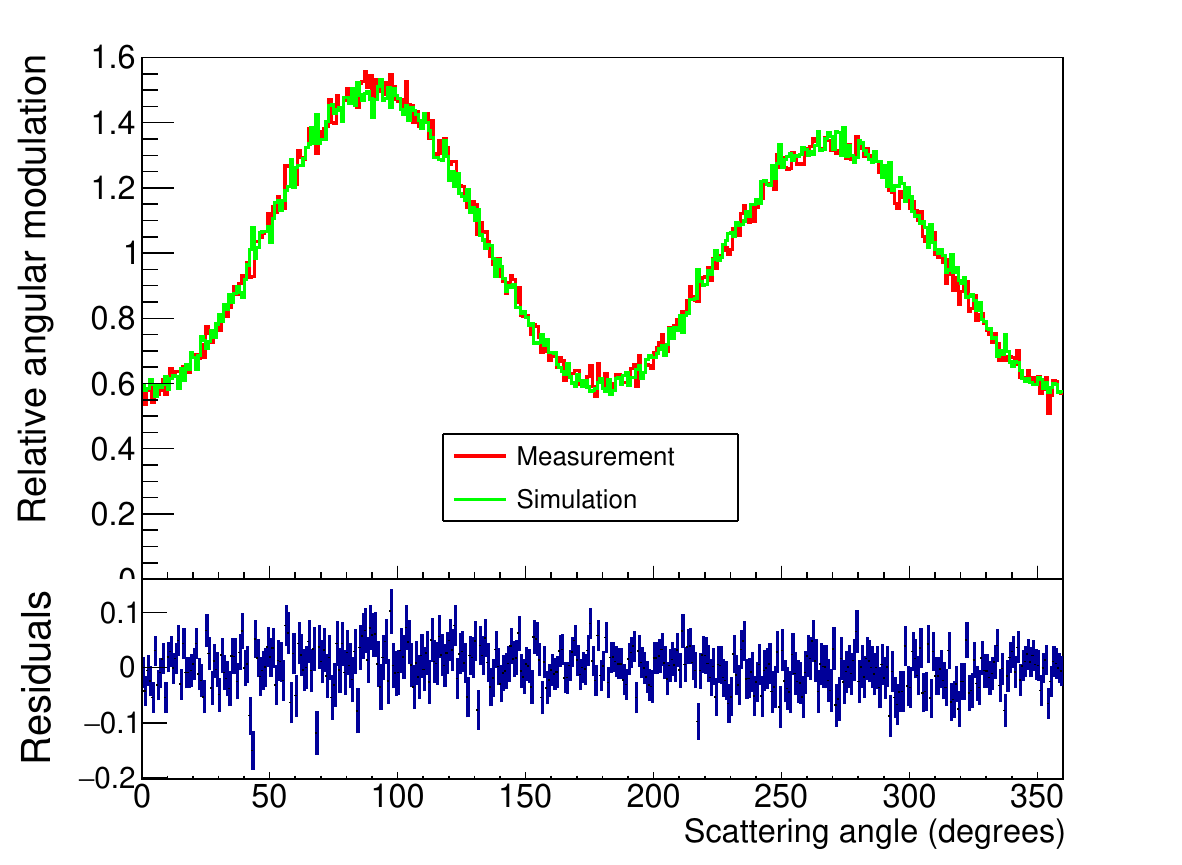}
    \includegraphics[width=.475\textwidth]{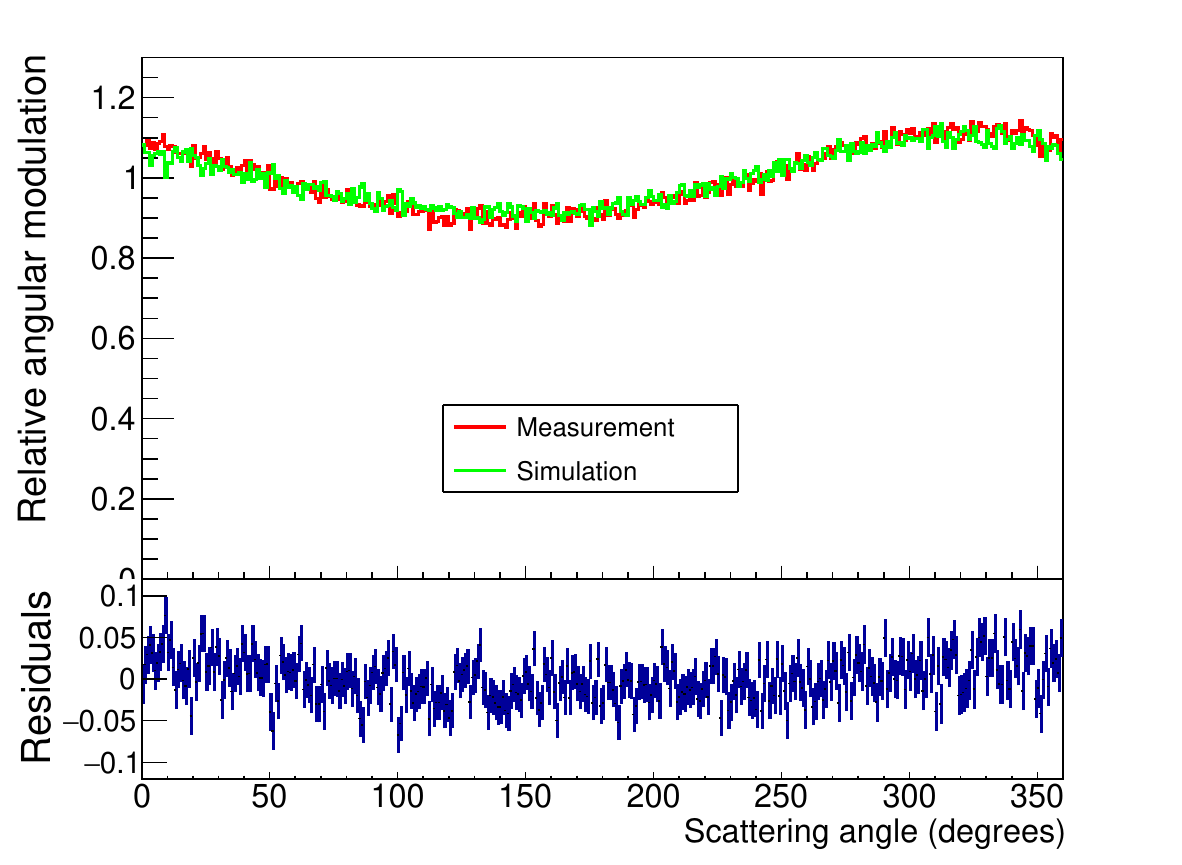}
    \caption{\label{Modulation curves - Esrange}Modulation curves for the measurement (red) and simulation (green) with a polarized beam (left) and unpolarized beam (right). Residuals (blue) are dominated by the alignment uncertainty in each setup. Curves have been normalized to the same area for visualization, but measured/simulated statistics are comparable. The presence of harmonics other than the 180$^{\circ}$ polarization signature is due to the beam hitting the scatterer off-center.}
\end{figure}

Resulting measured/simulated scattering-angle distributions for the polarized and the unpolarized beams are presented in Figure~\ref{Modulation curves - Esrange}. With only the simulated source position (x, y) as a free parameter, very good agreement is achieved in both cases, and no further fine-tuning is required, demonstrating the fidelity of the {\it XL-Calibur} Geant4 simulation. 

\section{Correcting for an offset beam}
\label{Correcting for an offset beam}
\subsection{Determining the beam offset}
The simulations can be used to evaluate the effects of the PSF migrating across the beryllium stick. Several independent effects contribute to such excursions: (1) pointing offset, (2) a possible (static) misalignment of the mirror, and (3) a (time-dependent) bending of the truss. These effects relate to the position of the focal point of the mirror and can be corrected for. Individual photons are further dispersed across the Be rod based on the mirror PSF and can only be corrected for on average.

The pointing offset (1) has an approximately linear relationship with the shift in the focal spot, with a $\sim$1.7~arcmin offset resulting in a 6~mm shift (the radius of the beryllium stick) in the focal plane of the detector\footnote{For large pointing offsets, a penalty in mirror effective area is incurred (vignetting)~\cite{Kamogawa.2022}. This effect is negligible for offsets within the beryllium stick.}. Since each recorded event is tagged with the pointing feedback from the Wallops Arc-Second Pointer~\cite{WASP}, individual events can  be corrected for the pointing offset. In the 2022 flight, a pointing stability of 10~arcsec (3$\sigma$) was achieved during observations of the Crab pulsar, which corresponds to a $\sim$0.6~mm jitter of the focal point.

A static misalignment (2) of the mirror can occur, e.g., due to mechanical alignment errors on ground. Although the mirror has no adjustable degrees of freedom during the flight, a systematic misalignment can partly be counteracted by adjusting the pointing in the opposite direction. The back-looking camera can be used to estimate the mirror alignment, under the assumption that the camera axis coincides with the mirror axis. To confirm the X-ray alignment, images from detector 17 are required. As shown in Section~\ref{Simulation and validation}, these can be sensitive to sub-mm offsets of the focal spot. A representative detector-17 image from the flight of {\it X-Calibur} can be found in~\cite{Abarr.2020p1n}.

\begin{figure}[hbtp]
    \centering
    \includegraphics[width=.85\textwidth]{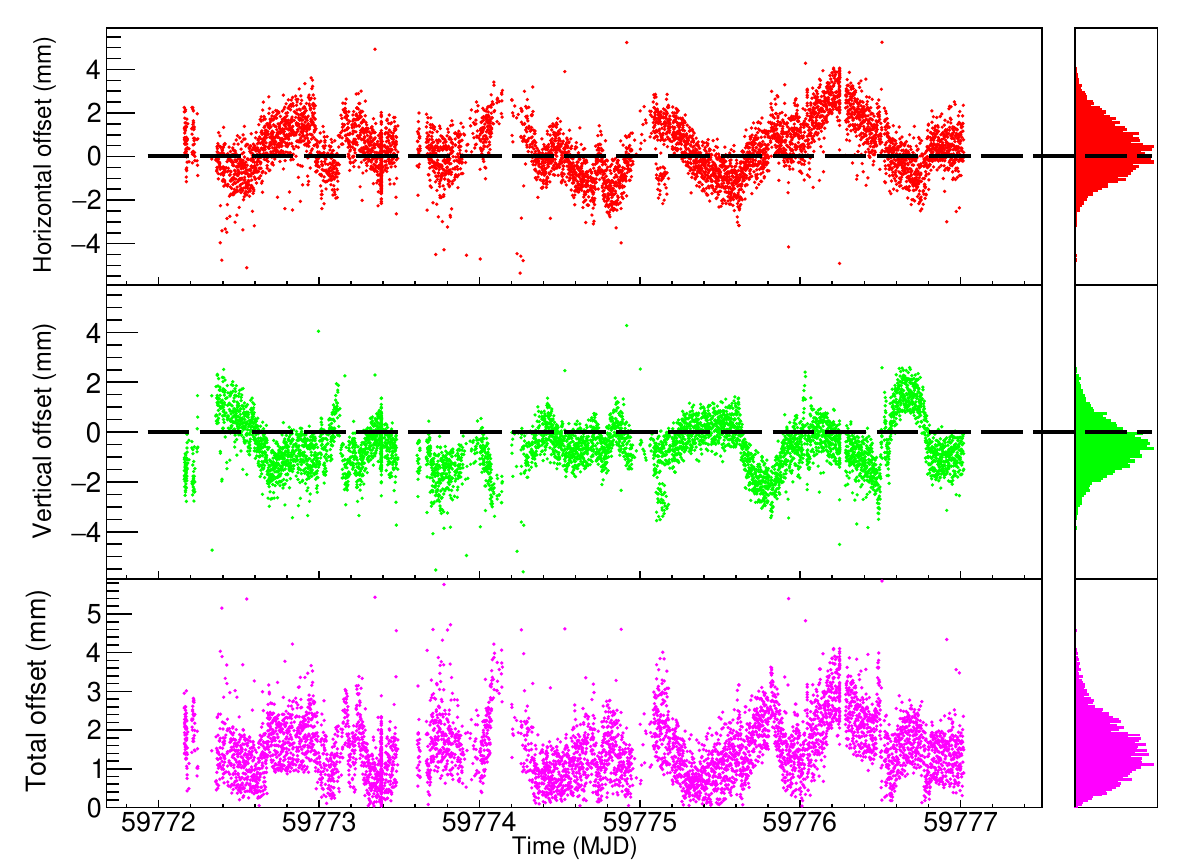}
    \caption{\label{Back-looking camera feedback}{\it XL-Calibur} truss alignment during the 2022 flight. Each data point corresponds to the center-point offset of the polarimeter, as reconstructed from the back-looking camera, for horizontal (red) and vertical (green) directions, and in total (magenta). The projections (right panel) reveal mean offsets of $\sim$0.3~mm and $\sim$0.6~mm in horizontal and vertical directions, respectively. Time-stamped data points from the back-looking camera allow event-by-event correction of the offset.}
\end{figure}

A bending of the truss (3) can arise, e.g., due to pointing-dependent effects of gravity on the 12-m long structure. To address this, an elevation-dependent sag correction (pointing offset) is implemented in the pointing system. Prior to flight, the shift in the optical axis of the back-looking camera is mapped out over the full range of elevations (up to $\sim$80$^{\circ}$). Figure~\ref{Back-looking camera feedback} shows the truss deflection during the 2022 {\it XL-Calibur} flight, as determined from the back-looking camera. 

Although the mean horizontal and vertical offsets are centered within $\sim$0.6~mm, excursions of up to $\sim$3-4 mm also occur. Integrating the image from detector 17 over a long time interval\footnote{For a Crab spectrum, approximately 10\% of the source photons from the mirror reach the bottom of the detector, whereby a 1~Hz source rate would result in $\sim$0.1~counts per second impinging on detector~17.} would destroy any temporal information. Since the back-looking camera can afford a high rate of offset data, scattering angles can be corrected essentially on an event-by-event basis. 

The instantaneous offset of the optical axis (back-looking camera) must be related to the most probable scattering site. This is taken as (mean~x, mean~y) in detector 17, calculated for events within the radius of the beryllium stick\footnote{Individual photons are further distributed around this point following the mirror PSF.}. A systematic simulation study of the offset in detector 17 (mean~x, mean~y) as a function of the true offset (x,~y) has been conducted for a grid of irradiation points across the beryllium stick. Both directions have a linear relation between the detector-17 offset and the true offset, with both regression coefficients exceeding 0.998. This is true for both the polarized and the unpolarized beam, indicating that (mean~x, mean~y) in detector 17 is a good proxy for the focal spot impinging on the top of the beryllium stick. Focal-spot excursions can thus be corrected on the time scale of the back-looking camera, instead of on the time scale of the integrated image in detector 17.

\subsection{Effects of a beam offset}
From any point along the central axis of the beryllium stick, the path length through the scatterer is the same for all azimuthal angles. Intrinsic differences between pixels, such as angular coverage, detection efficiency, threshold levels and noise characteristics are canceled out by the polarimeter rotation. The result for a fully centered beam is then a modulation curve as described by Eq.~(\ref{eqn:modcurve}), having only a 180$^{\circ}$ harmonic. 

\begin{figure}[hbtp]
    \centering
    \includegraphics[width=.95\textwidth]{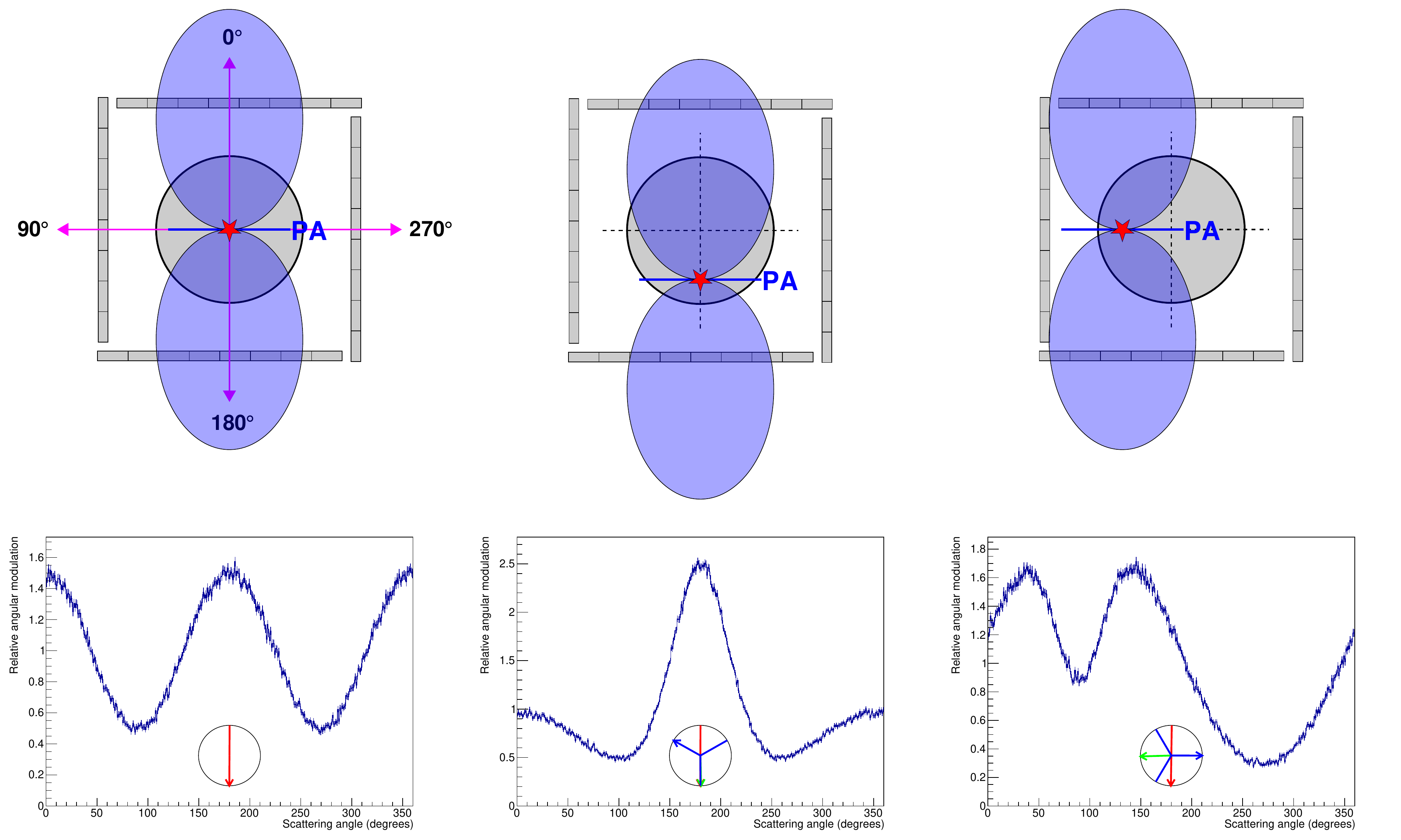}
    \caption{\label{Shifted peanuts in detector}Projected lobes (not to scale) of most probable scattering directions relative to the polarization angle \mbox{({\textquotedblleft}PA{\textquotedblright})} for scattering locations as indicated by the red stars (top), and simulated {\it XL-Calibur} modulation curves resulting from the corresponding beam offsets (bottom). The incident beam is infinitesimally narrow and has energy 53.3~keV. Red/green/blue arrows indicate phase for the 180$^{\circ}$, 360$^{\circ}$ and 120$^{\circ}$ components, respectively. Although the 180$^{\circ}$ phase is consistently reproduced (perpendicular to the polarization vector), the presence of other harmonics indicates that the 180$^{\circ}$ amplitude (giving the polarization fraction) is subject to systematic effects from an offset.}
\end{figure}

The assumption of an isotropic response ceases to be valid when the beam is offset from the center of the scattering element. Even for a fully centered beam, asymmetries can arise (see mirror PSF in Figure~\ref{PSF and D17}). A schematic view of how the offset alters the modulation curve is shown in Figure~\ref{Shifted peanuts in detector}.

Once the focal-point offset is known in the plane of the polarimeter, scattering angles can be calculated assuming this origin instead of the center of the scatterer, while the absorption site is assigned a randomized position within the triggered pixel\footnote{The nominal angular acceptance in azimuth of a pixel is $\sim$11.25$^{\circ}$ (360$^{\circ}$ subtended by 32 pixels, as shown in Figure~\ref{Shifted peanuts in detector}). Depending on the relative positions of the scattering site and absorbing pixel, the angular acceptance can vary from $\sim$5$^{\circ}$ to $\sim$30$^{\circ}$.}.
This correction only takes the azimuthal-angle acceptance (in the plane of the detector) into account. The change in polar-angle acceptance (axial direction) cannot be accounted for, as the height of the interaction along the axis of the beryllium stick is not known. For the same reason, the difference in path length through the stick cannot be treated. This effect is small compared to the change in angular acceptance caused by the beam offset: for a 30 keV photon, shifting the interaction site by 3 mm from the center (half of the stick radius) changes the attenuation in the perpendicular direction to the nearest wall by less than 10\%, while the azimuthal-angle acceptance in the same direction changes by more than 40\%. For higher energies, the relative difference becomes even more pronounced, as the absorption probability decreases while the solid angle remains unchanged.

\subsection{Systematic error with and without offset correction}
\begin{figure}[hbtp]
    \centering
    \includegraphics[width=.475\textwidth]{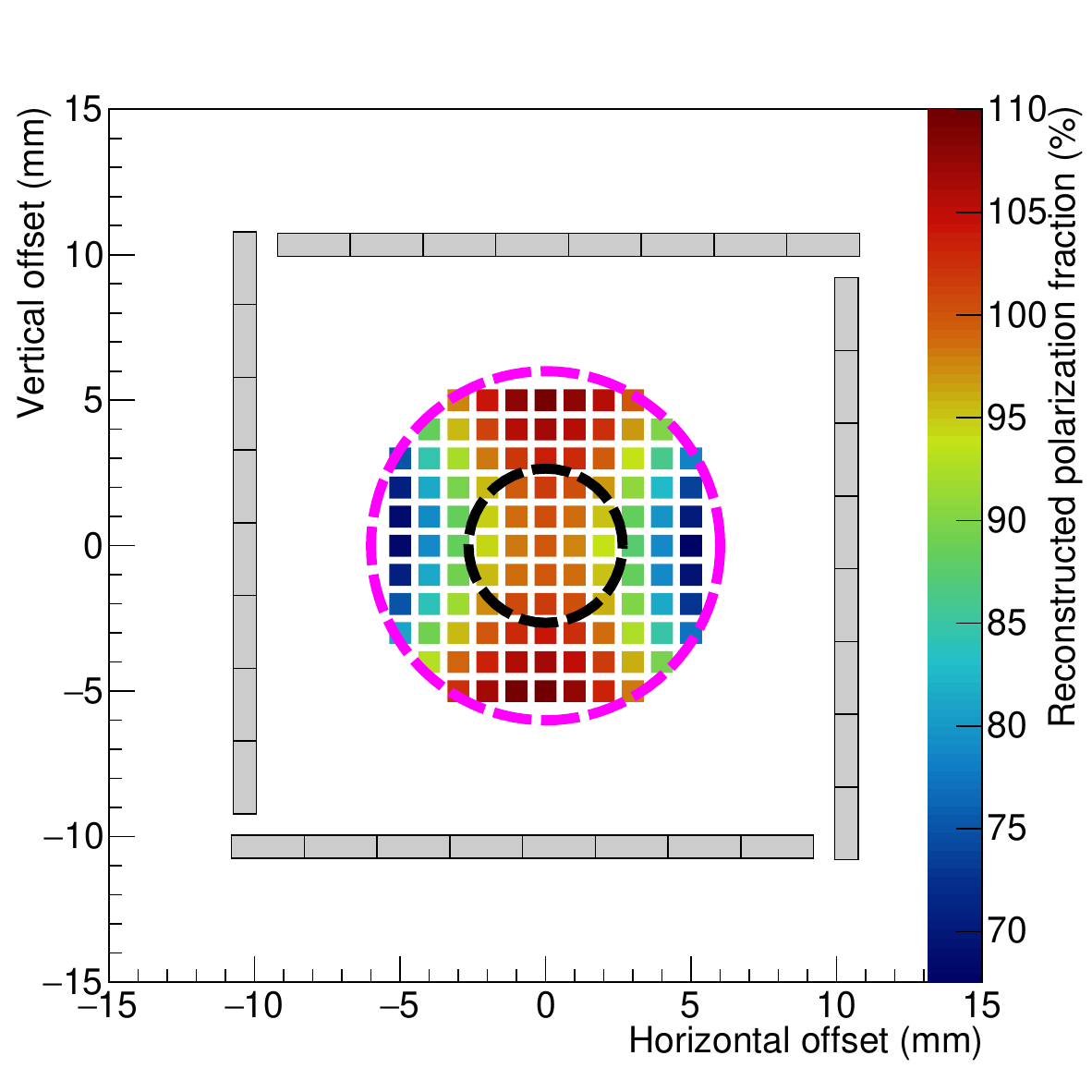}
    \includegraphics[width=.475\textwidth]{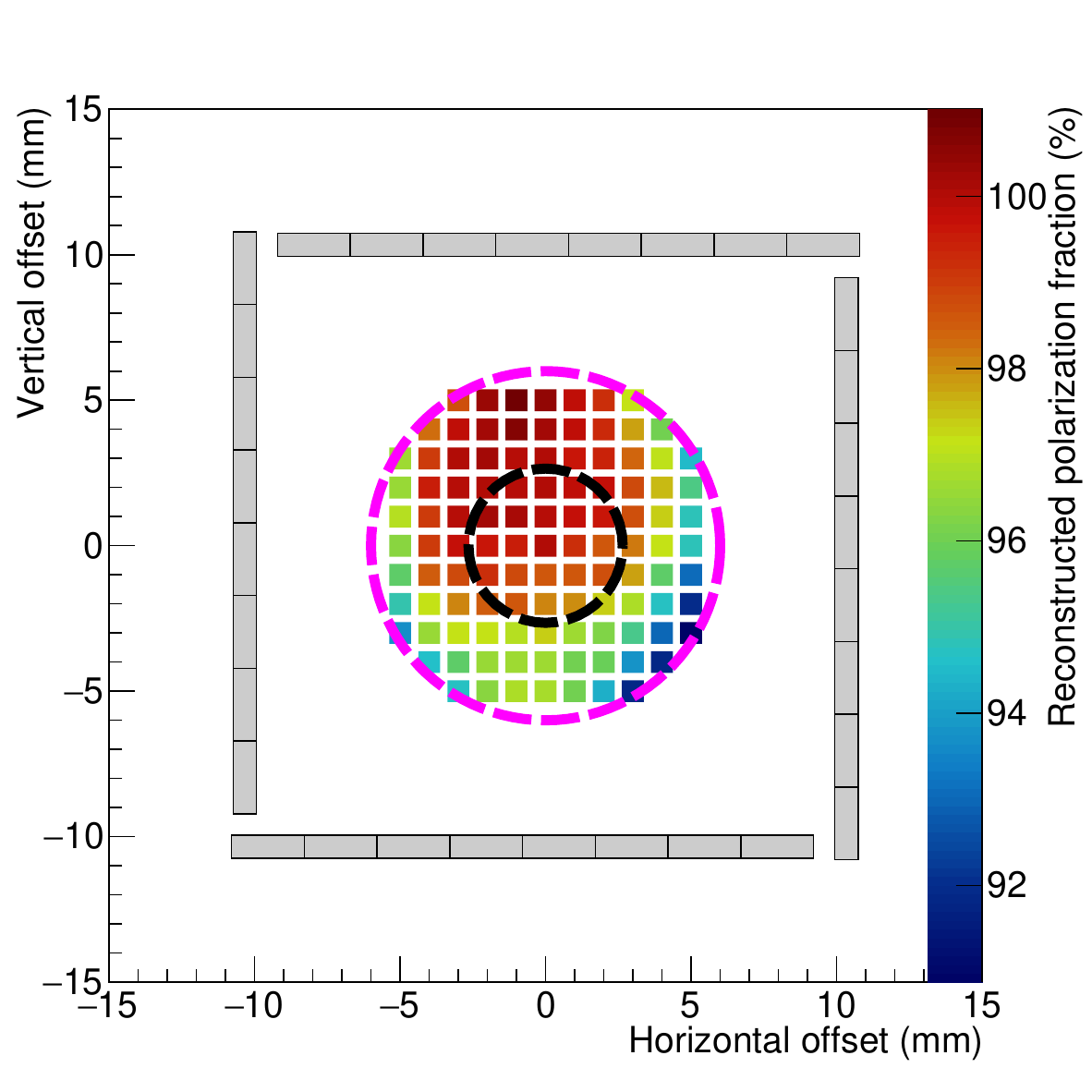}
    \caption{\label{Grid scan polarized}Reconstructed polarization fraction for an offset mirror beam with 100\% polarization (horizontal direction), in the absence of offset correction (left) and with a correction based on calculating scattering angles using detector-17 (mean~x, mean~y) (right). The color scale is normalized relative to the central point (zero beam offset, corresponding to the $\sim$0.3~mm shift shown in Figure~\ref{PSF and D17}). Irradiation points where the PSF is centered outside the beryllium stick (magenta circle) can also be kept, and benefit even more from the offset correction. Such points have been excluded here, as the flux drops to below 50\% once the focal point exceeds the boundary of the beryllium stick. During the 2022 flight, 90\% of the observation time was spent within a total offset of less than 2.65~mm (black dashed circle). The error remaining after correction (right) is slightly shifted upwards, likely resulting from the mirror asymmetry which shows a bias in the zenith direction (see Figure~\ref{PSF and D17}).}
\end{figure}

\begin{figure}[hbtp]
    \centering
    \includegraphics[width=.475\textwidth]{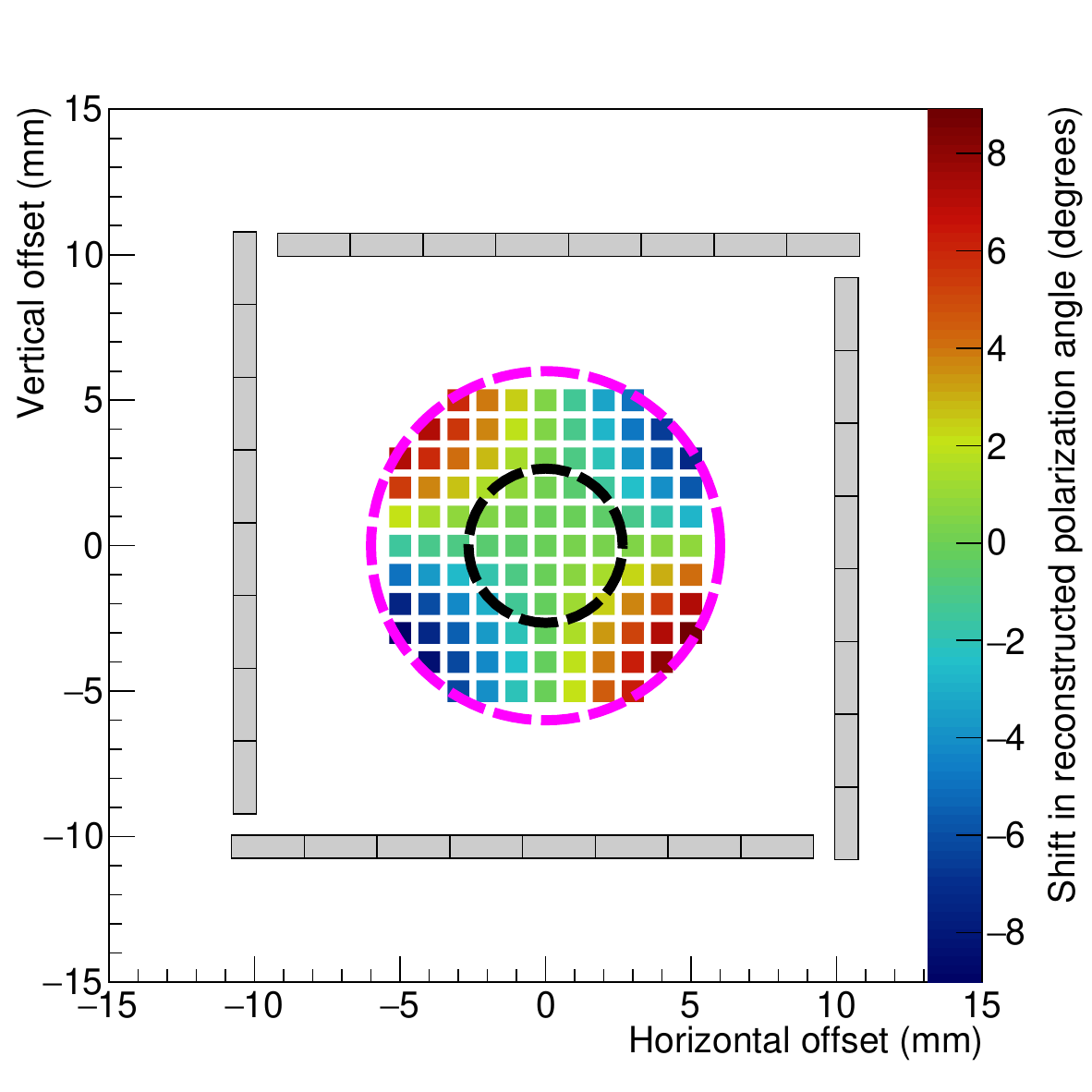}
    \includegraphics[width=.475\textwidth]{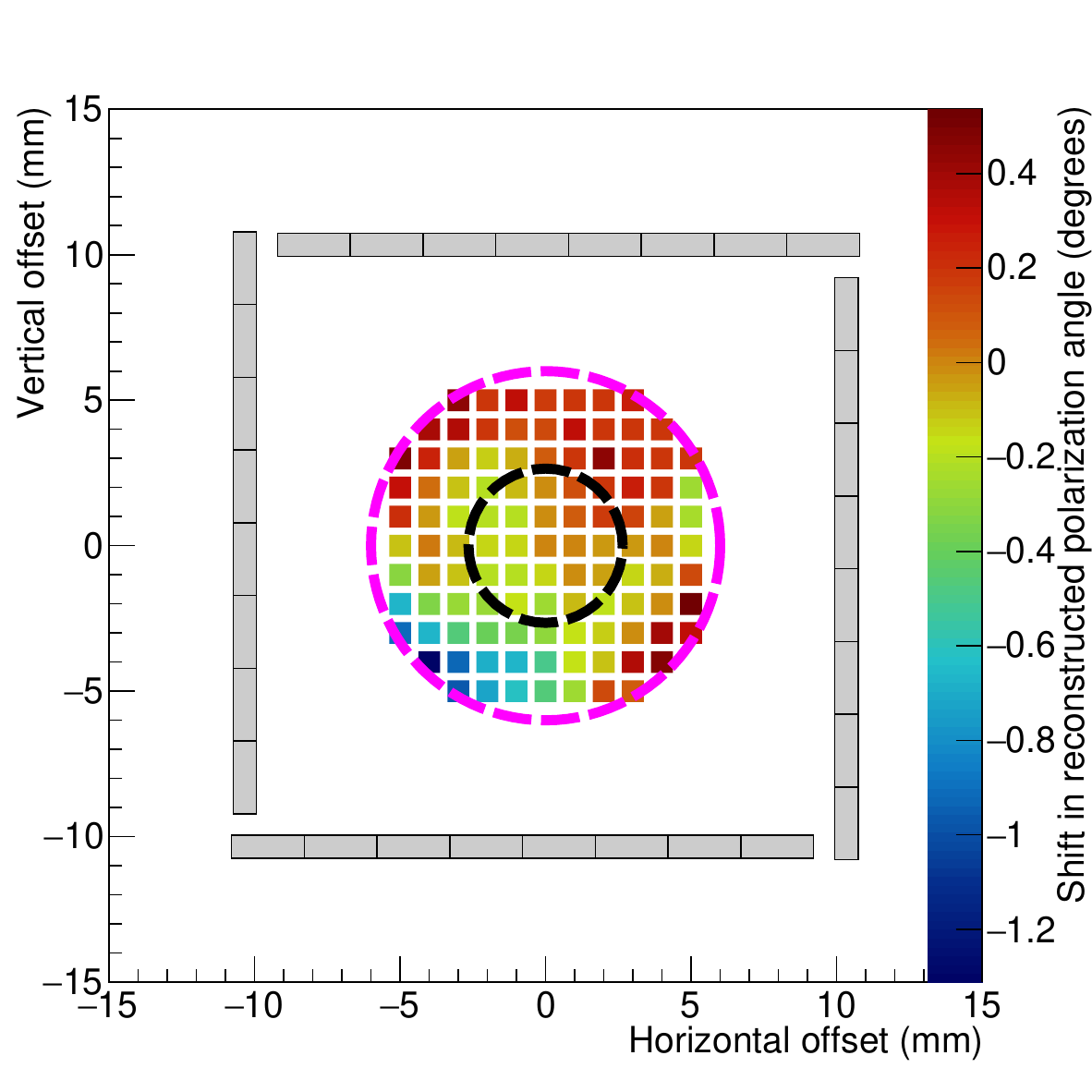}
    \caption{\label{Grid scan - polarization angle - polarized}Shift in reconstructed polarization angle resulting from an offset of the incident mirror beam with 100\% polarization (horizontal direction), in the absence of offset correction (left) and with a correction based on calculating scattering angles using detector-17 (mean~x, mean~y) (right). The color scale is defined relative to the central point (zero beam offset, corresponding to the $\sim$0.3~mm shift shown in Figure~\ref{PSF and D17}. Diagonal offsets are more affected than those perpendicular to or parallel with the polarization vector. After correction, systematic effects are small, revealing fluctuations due to the structure of the mirror PSF. During the 2022 flight, 90\% of the observation time was spent within a total offset of less than 2.65~mm (black dashed circle).}
\end{figure}

The detector-17 offset (mean~x, mean~y), calculated within the 6~mm radius of the scatterer, is assumed as the scattering vertex for the offset correction\footnote{If the focal point is close to the edge of the scatterer, some photons from the mirror can interact outside this radius, thus (mean~x, mean~y) calculated within the stick will always be less than the offset of the PSF.}. 
Although the correction is not perfect on an event-by-event basis -- due to incomplete information about the scattering kinematics of individual events -- it is highly successful in reducing the systematic error on the reconstructed polarization resulting from a beam offset. Figure~\ref{Grid scan polarized} shows the effect on the reconstructed polarization fraction for the polarized case, with and without offset correction based on the image from detector 17, for a grid of hypothetical beam offsets. Effects on the reconstructed polarization angle are shown in Figure~\ref{Grid scan - polarization angle - polarized}.

In the absence of a correction, systematic errors of up to several tens of percentage points in the reconstructed polarization fraction can arise for offsets within the radius of the scattering element. Even modest offsets of 1~mm (2~mm) result in absolute errors of $>$2~percentage points ($>$6~percentage points). For a polarized beam, shifts along the polarization vector result in a reduction in apparent polarization, while perpendicular shifts increase the reconstructed polarization relative to the central point. Nonphysical values with apparent polarization fraction in excess of 100\% can thus occur. After correction, systematic errors are significantly reduced, e.g., even a 2~mm offset now gives $<$2~percentage points of error, for any offset direction. The knowledge of the location of the focal spot is thus essential for high-accuracy polarization measurements. 

The reconstructed polarization angle is similarly altered in the absence of a correction, and both the magnitude and the sign of the systematic error depend on the direction of the beam offset. The largest error is incurred when the offset is diagonal relative to the polarization direction, reaching almost a $\pm$10$^{\circ}$ shift for a 6~mm offset. Offsets parallel with or perpendicular to the polarization vector are shifted by less than $\pm$2$^{\circ}$ for the same offset. With offset correction, the true polarization direction is recovered within $\sim$1.5$^{\circ}$ for any systematic offset on the beryllium stick.

\begin{figure}[hbtp]
    \centering
    \includegraphics[width=.475\textwidth]{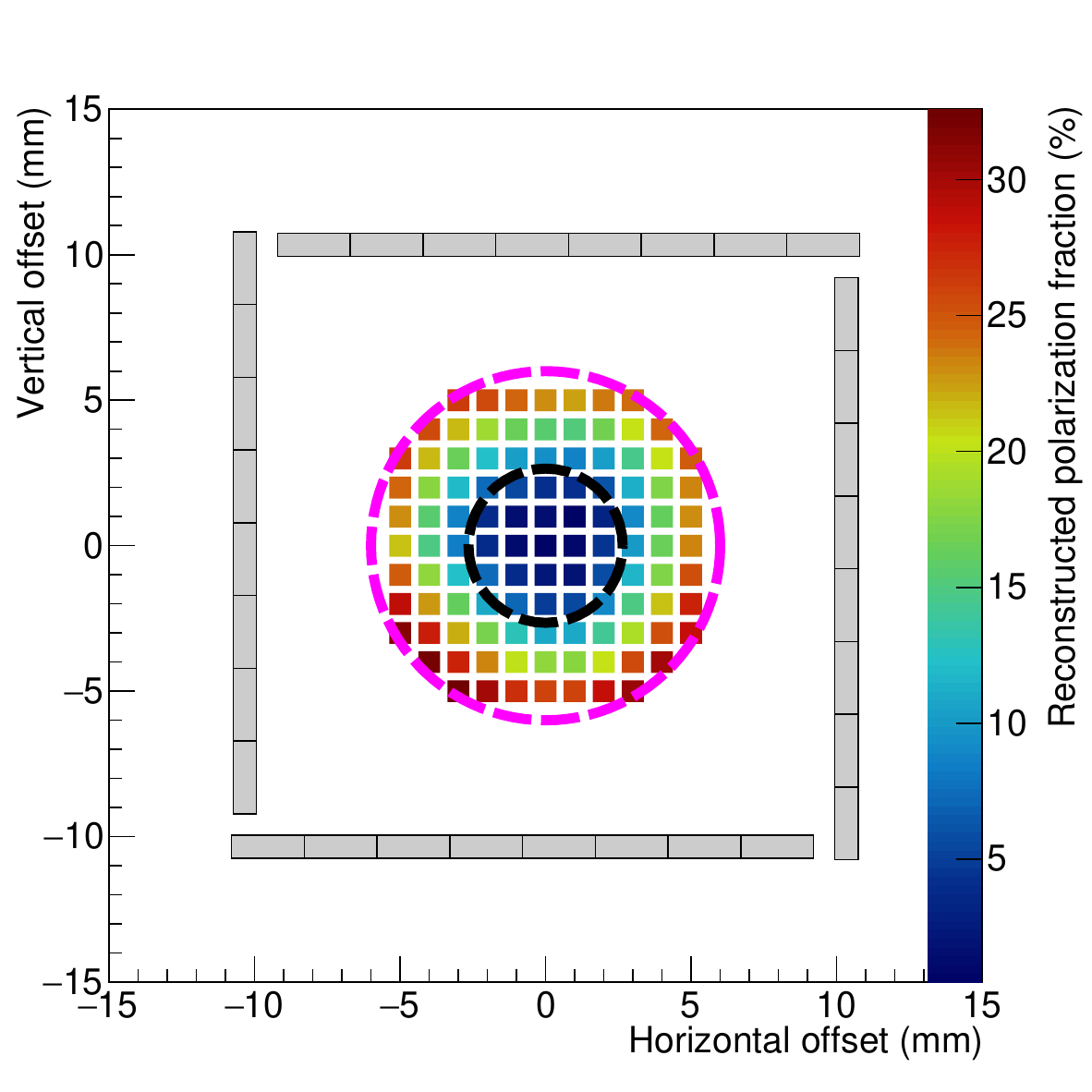}
    \includegraphics[width=.475\textwidth]{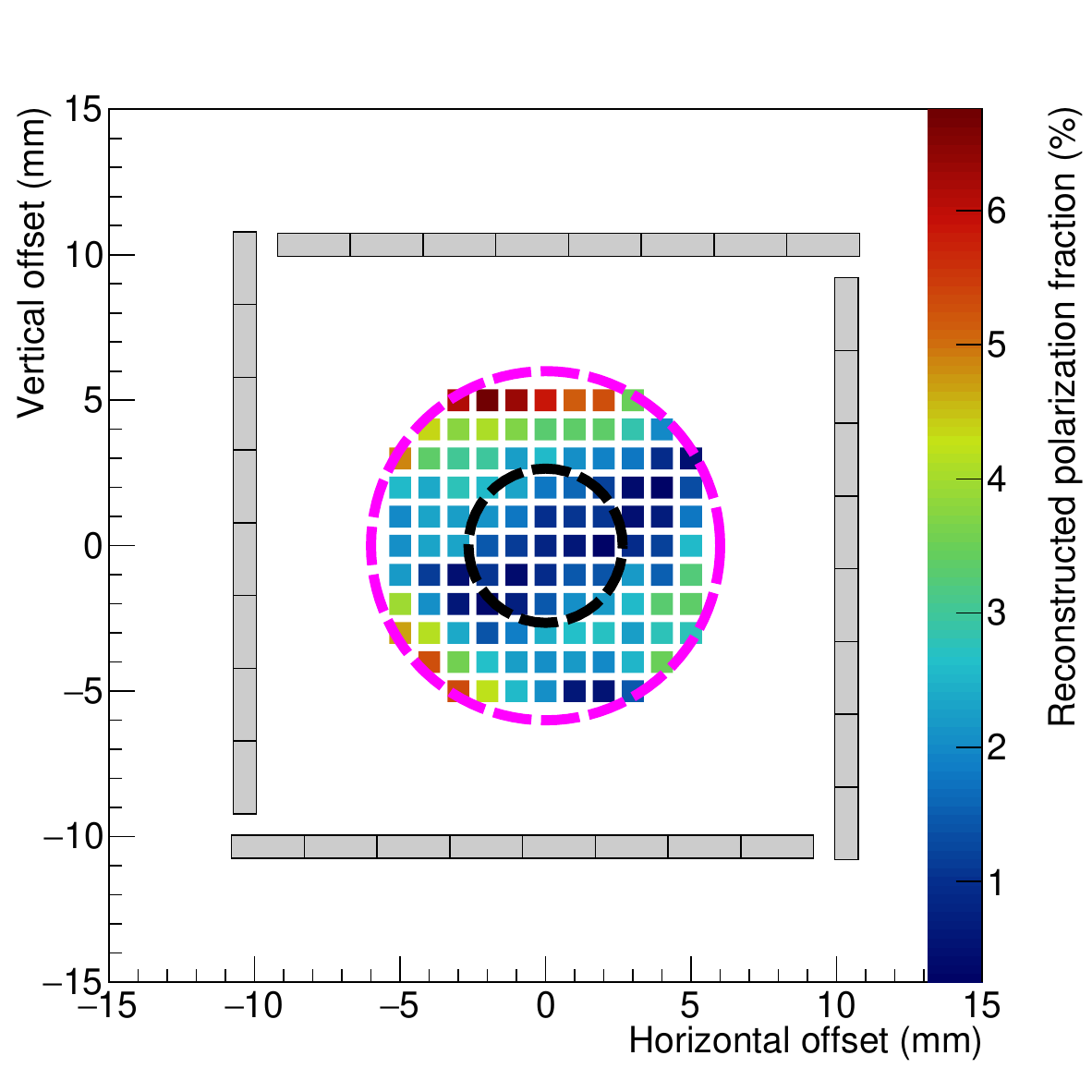}
    \caption{\label{Grid scan unpolarized}Spurious polarization resulting from an offset of the incident mirror beam in the unpolarized case, in the absence of offset correction (left) and with a correction based on calculating scattering angles using detector-17 (mean~x, mean~y) (right). The color scale is normalized relative to the central point (zero beam offset, corresponding to the $\sim$0.3~mm shift shown in Figure~\ref{PSF and D17}) for the 100\% polarized beam. During the 2022 flight, 90\% of the observation time was spent within a total offset of less than 2.65~mm (black dashed circle).}
\end{figure}

Results for an unpolarized beam are shown in Figure~\ref{Grid scan unpolarized}. In the absence of an offset correction, even a 2-mm offset can cause spurious polarization as high as 5\%, and the effect increases drastically for larger offsets. This can immediately preclude the sought percent-level MDP, and underlines the importance of applying an offset correction. A systematic error from offset in one direction is not obviously counteracted by an error caused by an offset in the opposite direction: instead, these errors may compound. Once a correction is applied, systematic errors become significantly lower\footnote{Here, the impact of the bias due to the positive-definite nature of the measurement~\cite{Mikhalev_Pitfalls_2018} is restricted by high statistics in each simulated grid point -- approximately an order of magnitude more than what is expected for a source observation during the balloon flight.}. The remaining asymmetry in the corrected grid results from the anisotropy in the beam PSF, seen in Figure~\ref{PSF and D17}.

\section{Observation-specific modulation response and spectro-polarimetry}
\label{Modulation response and spectro-polarimetry}
\begin{figure}[hbtp]
    \centering
    \includegraphics[width=.7\textwidth]{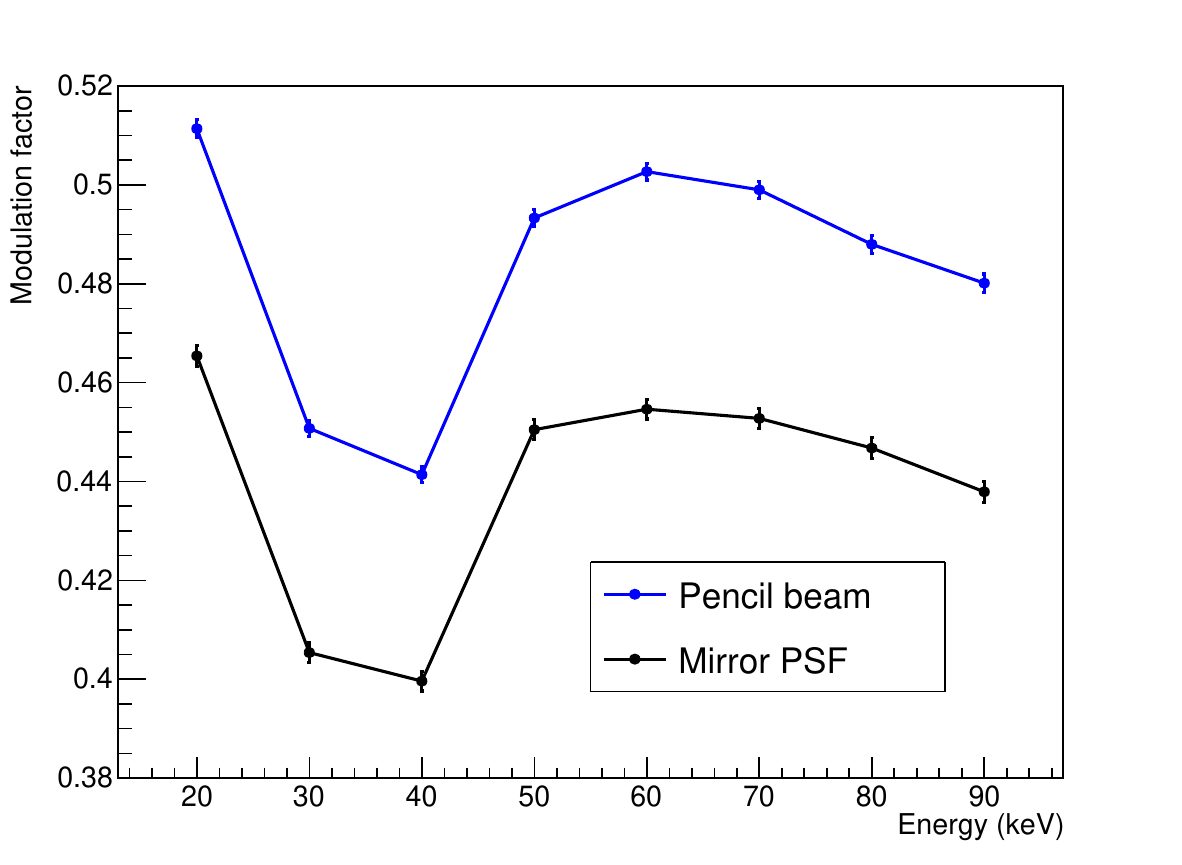}
    \caption{\label{Modulation and energy}{\it XL-Calibur} modulation response to a 100\% polarized flux ($\mu_{100}$) for a pencil beam (blue) and the simulated mirror PSF (black). The dip around 30-40~keV is caused by characteristic (K$_{\alpha}$) X-ray emission from the cadmium in the CZT detectors. Such secondary photons carry no polarimetric information pertaining to the original flux, and if they interact in a second pixel while the primary interaction is below the detection threshold, they will appear as a primary event and cannot be rejected, thus resulting in a suppression of the polarization detected for a given flux.}
\end{figure}

For accurate polarization measurements, the modulation response ($\mu_{100}$) must be determined for a given observation. In the simulation, factors such as float altitude, source spectrum, elevation-dependent atmospheric attenuation and systematic pointing offset can be taken into account. Additionally, for spectro-polarimetry~\cite{Kislat_XrayHandbook}, the energy dependence of the modulation response must be considered. This is shown in Figure~\ref{Modulation and energy}, comparing results for an infinitesimally narrow pencil beam and for the simulated mirror PSF. Within the {\it XL-Calibur} energy range, $\mu_{100}$ can vary by as much as $\sim$7~percentage points. The mirror PSF reduces the value by an additional $\sim$5~percentage points as compared to the pencil beam.

\begin{figure}[hbtp]
    \centering
    \includegraphics[width=.7\textwidth]{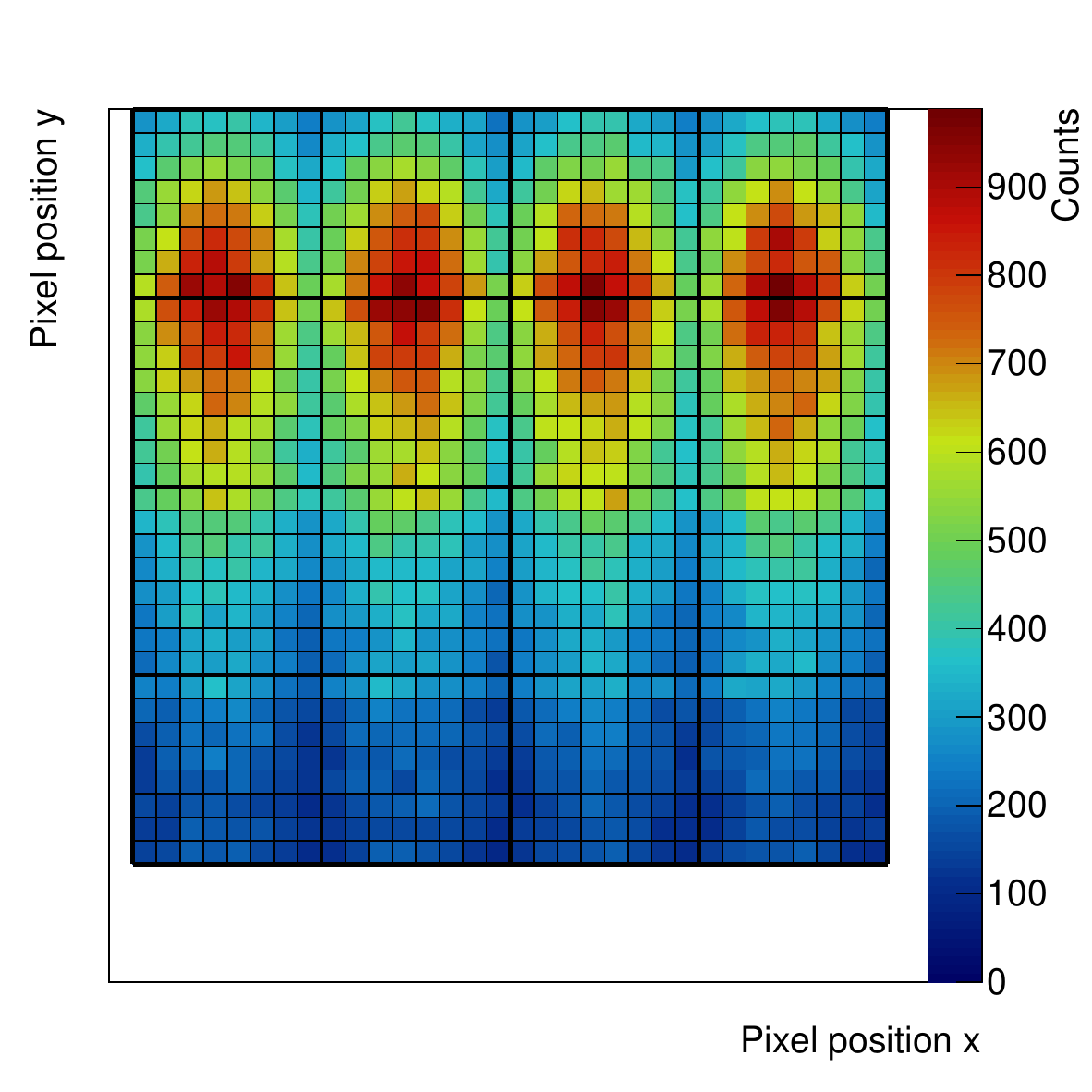}
    \caption{\label{Pixel heat map}Pixel hit-map of the 16 CZTs for a simulated Crab observation. The four vertical regions (delimited by thick black lines) correspond to the detector walls. Horizontal regions indicate the four CZT \mbox{{\textquotedblleft}rings{\textquotedblright}} along the length of the beryllium stick. Lobes with high count intensity arise from the solid-angle effect: as seen from the scatterer, pixels near the center of a wall subtend a larger solid angle than corner pixels. The beam from the mirror is incident \mbox{{\textquotedblleft}from the top{\textquotedblright}} of the image. The vertical position of the lobes correlates to the beam energy (higher beam energies penetrate deeper into the beryllium stick), and the map thus provides indirect information on the polar scattering angles.}
\end{figure}

When populating the modulation curve, individual event weights can provide observation-specific corrections such as for a systematic offset. The weighting strategy is based on the expression for the minimum detectable polarization~\cite{Weisskopf_MDP_2010}, which can be expressed at 99\% confidence level as
\begin{equation}
    \text{MDP}=\frac{4.29}{\mu_{100}S}\sqrt{N},
    \label{MDP equation}
\end{equation} 
where $N=S+B$ is the total number of signal ($S$) and background ($B$) counts. Following Eq.~(\ref{MDP equation}), two types of weights will be considered: the modulation factor, $\mu$, and the number of signal events, incorporated in the factor $N$. Maximizing $\mu$ reduces the uncertainty on the reconstructed polarization fraction (see, e.g,~\cite{Mikhalev_Pitfalls_2018} for a discussion), while the benefit in weighting for $N$ comes from optimizing the signal-to-background ratio as illustrated in Figure~\ref{Pixel heat map}.

The pixel hit-map for a source observation is quantitatively different from a background observation, which is expected to be uniformly populated. Applying event weights based on the relative number of counts in a given pixel can thus favor signal events from this topology, while suppressing background. For example, events near the bottom of the CZT walls have a low probability of being signal and should be assigned low weights. Based on Eq.~(\ref{MDP equation}), these weights are applied as $\sqrt{N}$.

\begin{figure}[hbtp]
    \centering
    \includegraphics[width=.7\textwidth]{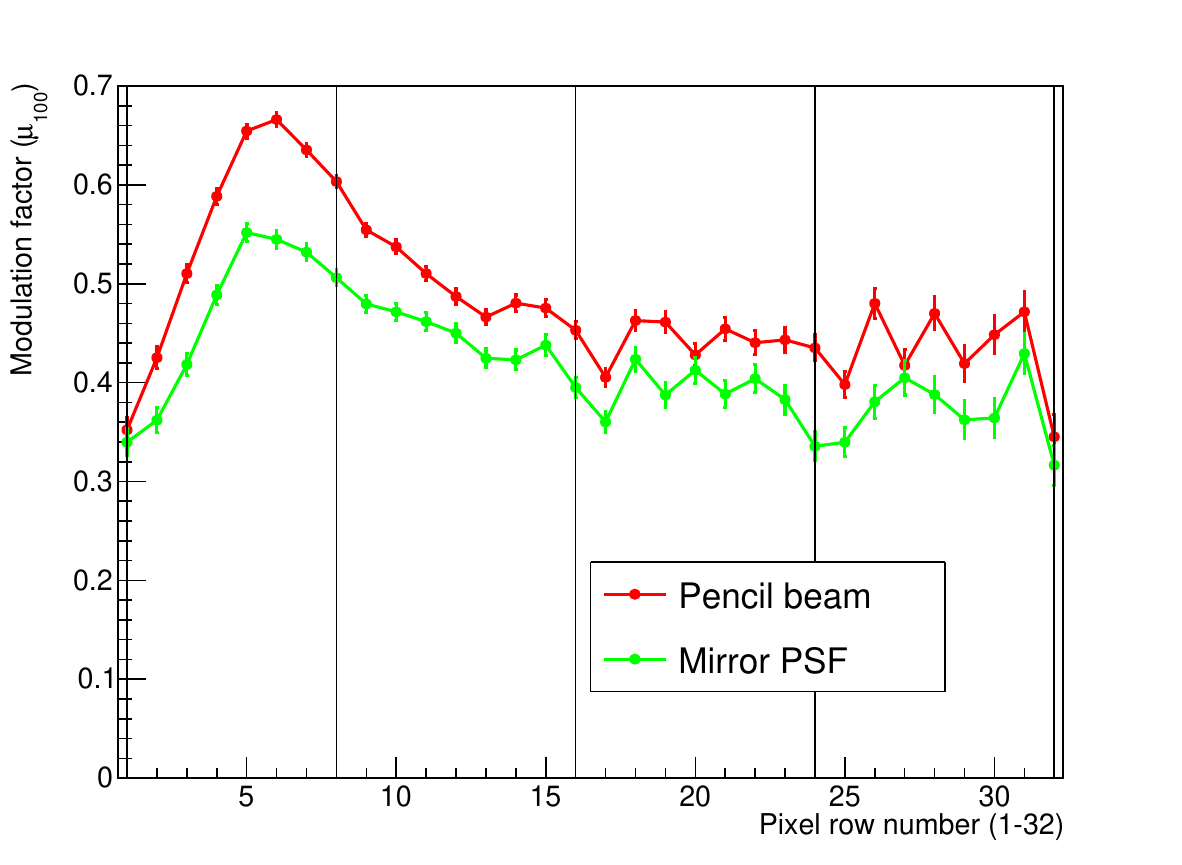}
    \caption{\label{Row-dependent M100}Simulated modulation factor for each pixel row of the {\it XL-Calibur} detector for a Crab spectrum, for a pencil beam (red) and the mirror PSF (green). Vertical lines indicate the four CZT rings, and dips at rows 17 and 25 result from the gaps between adjacent rings. The top of the beryllium stick is approximately centered on the first ring, where the highest values of $\mu_{100}$ are seen, resulting from scattering at close to 90$^{\circ}$ in polar angle. The increase in error-bar size at high row numbers results from the reduced number of polarization events reaching the bottom of the detector (see Figure~\ref{Pixel heat map}).}
\end{figure}

Following the Klein-Nishina formula, Eq.~(\ref{eqn:klein-nishina}), the amplitude of the azimuthal modulation for a polarized flux depends on the polar scattering angle. This is encoded in the modulation factor, and photons scattering at polar angle $\sim$90$^{\circ}$ are most favorable for polarimetry. Figure~\ref{Row-dependent M100} shows this effect for {\it XL-Calibur}, through the modulation response for each row of CZT pixels in the detector.

For some pixel rows, modulation factors as high as $\sim$0.67 ($\sim$0.55) are seen for the pencil (mirror PSF) beam. Folding this modulation response with the pixel-hit distribution (Figure~\ref{Pixel heat map}) gives the effective modulation factor for a given observation. Instead of using the overall modulation factor for determining the polarization fraction of the flux, individual scattering events can be weighted with their corresponding modulation factor, following Figure~\ref{Row-dependent M100}. Following Eq.~(\ref{MDP equation}), the $\mu$ weights apply linearly.

An optimized analysis~\cite{Kislat_XrayHandbook} of flight data can benefit from event weights based on the product $\mu\sqrt{N}$, through the corresponding reduction in MDP. Within the established simulation framework, both $\mu$ and $\sqrt{N}$ weighting schemes (Figure~\ref{Pixel heat map}, Figure~\ref{Row-dependent M100}, respectively) can be generated for any systematic offset of the mirror PSF. Thus, even if a systematic offset is present in a measurement, the analysis can benefit from the correction described here.

\section{Summary and conclusions}
\label{Summary and conclusions}
For a polarimeter at the focus of an X-ray mirror, shifts in the focal point can occur due to pointing offsets, possible mirror misalignment and/or elevation-dependent structural deflection. In the absence of an offset correction, large systematic errors are incurred the reconstructed polarization parameters, both for polarized and for unpolarized beams. Effects for a mirror PSF, as described here, are not as severe as for a highly collimated synchrotron beam~\cite{Beilicke.2014}. This is because the extended PSF causes some photons to scatter at lower offsets, reducing the systematic error, while photons incident at larger offsets (leading to high systematic error) instead have a probability of missing the stick. Corrections can be applied by calculating the Compton-scattering angle not from the symmetry axis of the stick, but from the most probable scattering location at any given time. This is allowed by combining information from an imaging CZT, located underneath the beryllium scatterer, and a back-looking camera mounted on the optical axis of the mirror, to offer offset corrections on an event-by-event level.

Based on a simulation bench-marked with laboratory testing using radioactive sources, the polarimeter response for flight can be studied. Estimated values of $\mu_{100}$ for the two primary targets for the upcoming flight from Esrange, taking into account source spectra, atmospheric attenuation, mirror response~\cite{Kamogawa.2022}, energy dependence (Figure~\ref{Modulation and energy}) and detector response, are shown in Table~\ref{M_100 table}. Assuming a modulation factor \mbox{$\mu_{100} \approx 0.5$}~\cite{Abarr.2021}, derived from earlier simulations based on synchrotron beams~\cite{Beilicke.2014}, would then systematically under-estimate reconstructed polarization fractions.

\begin{table}[H]
\centering
\caption{\label{M_100 table}Representative $\mu_{100}$ values for primary northern-hemisphere targets characterized by photon index $\Gamma$. Final values will be calculated accounting for the float altitude achieved and source variability.}
    \begin{tabular}{cc}
    \hline\hline
    Source & $\mu_{100}$ \\
    \hline
    Crab $(\Gamma=2.11)$ & $(42.6 \pm 0.2)\%$\\
    Cygnus X-1 low-hard state $(\Gamma=1.70)$ & $(43.1 \pm 0.2)\%$\\
    Cygnus X-1 high-soft state $(\Gamma=2.47)$ & $(43.0 \pm 0.2)\%$\\
    \hline
    \end{tabular}
\end{table}

The minimum-detectable polarization can be improved through event weighting, using a factor $\mu\sqrt{N}$. These weights address the fact that detector rows subtend different polar-angle ranges. The benefit of the weighting depends on the source spectrum (through $\mu$) as well as on the signal-to-background scenario in flight (through $\sqrt{N}$). Based on simulations with a conservative 1:1 ratio of signal and background\footnote{The Crab signal rate is expected from simulations to exceed 1~Hz~\cite{Abarr.2021}. During the 2022 flight, a background rate of $\sim$0.5~Hz was measured~\cite{Iyer.2023}.}, the MDP for a 1~Crab source at $\sim$45$^{\circ}$ elevation is expected to improve from $\sim$4\% to $\sim$3\% for an on-source integration time of 24~hours.

{\it XL-Calibur} is scheduled for a flight from Esrange Space Center, Sweden, in 2024. With the percent-level MDP achievable, the instrument is expected to provide strong polarimetric constraints for astrophysical sources in the hard X-ray range, $\sim$15--80~keV range, complimentary to the soft X-ray regime, 2-8~keV, accessible to the Imaging X-Ray Polarimetry Explorer (IXPE)~\cite{IXPE_Weisskopf}. Joint observations are planned for the upcoming flight, to best utilize synergies between the measurements for polarimetry and spectro-polarimetry.

\section*{Acknowledgments}
{\it XL-Calibur} is funded in the US by the NASA APRA (Astrophysics Research and Analysis) program. Members acknowledge funding by grants 80NSSC20K0329 and 80NSSC24K0205. 

KTH authors acknowledge funding from the Swedish National Space Agency (Rymdstyrelsen), grants 2020-00201, 2022-00178, and from the Swedish Research Council (Vetenskapsr{\aa}det), grant 2021-05128. NI thanks the KTH Space Center for funding support. 

The X-ray mirror was characterized at SPring-8 using the beam time of 2019B1221, 2020A0746, 2020A1214, 2020A1298, 2022B1255, and 2023A1469. HM and HT are supported by JSPS KAKENHI Grant Numbers JP19H05609, JP19H01908, JP20H00175 and JP23H00128, and by the ISAS program for small-scale projects.

{\it XL-Calibur} shared integration space with the {\it Sunrise-III} mission at Esrange, and we thank them for the practical arrangements during calibration tests of the polarimeter.

\bibliographystyle{ieeetr}
\bibliography{XL_response}

%
%
\end{document}